\documentclass[prd,preprint,nofootinbib,superscriptaddress,showpacs]{revtex4}

\usepackage{amsmath,amssymb}
\usepackage{graphicx}
\usepackage{dcolumn}
\usepackage{bm}

\newcommand{\nc}{\newcommand}  

\nc{\beq}{\begin{equation}}  
\nc{\eeq}{\end{equation}}  
\nc{\beqa}{\begin{eqnarray}}  
\nc{\eeqa}{\end{eqnarray}}  
\nc{\bea}{\begin{eqnarray}}  
\nc{\eea}{\end{eqnarray}}  
\nc{\ra}{\rightarrow}  
\nc{\lsim}{\begin{array}{c}\,\sim\vspace{-21pt}\\< \end{array}}  
\nc{\gsim}{\begin{array}{c}\sim\vspace{-21pt}\\> \end{array}}  

\nc{\MM}{\ensuremath{\mathcal{M}}}  

\begin{document}
\title{Electroweak constraints on warped models with
custodial symmetry.}

\date{\today}

\vspace*{-2.3cm}  
\begin{flushright}  
\normalsize{  
ANL-HEP-PR-07-02\\
hep-ph/0701055  
  }  
\end{flushright}  
%


\author{Marcela Carena}
\affiliation{ Fermi National Accelerator
Laboratory, P.O. Box 500,  Batavia, IL 60510, USA}
\author{Eduardo Pont\'on}
\affiliation{Department of Physics, Columbia University,\\
             538 W. 120th St, New York, NY 10027, USA} 
\author{Jos\'e Santiago}
\affiliation{ Fermi National Accelerator
Laboratory, P.O. Box 500,  Batavia, IL 60510, USA}
\author{C.E.M. Wagner}
\affiliation{HEP Division, Argonne National Laboratory,  
9700 Cass Ave., Argonne, IL 60439, USA}  
\affiliation{Enrico Fermi Institute and Kavli Institute for
  Cosmological Physics, \\ 
   Univ. of Chicago, 5640  
Ellis Ave., Chicago, IL 60637, USA}

\begin{abstract}
It has been recently argued that realistic models with warped extra
dimensions can have Kaluza-Klein particles accessible at the Large
Hadron Collider if a custodial symmetry, $SU(2)_V\times P_{LR}$, is
used to protect the $T$ parameter and the coupling of the left-handed
bottom quark to the $Z$ gauge boson.  In this article we emphasize
that such a symmetry implies that the loop corrections to both the $T$
parameter and the $Zb_L \bar{b}_L$ coupling are calculable.  In
general, these corrections are correlated, can be sizable, and should
be considered to determine the allowed parameter space region in
models with warped extra dimensions and custodial symmetry, including
Randall-Sundrum models with a fundamental Higgs, models of gauge-Higgs
unification and Higgsless models.  As an example, we derive the
constraints that arise on a representative model of gauge-Higgs
unification from a global fit to the precision electroweak
observables.  A scan over the parameter space typically leads to a
lower bound on the Kaluza-Klein excitations of the gauge bosons of
about $2-3$ TeV, depending on the configuration.  In the fermionic
sector one can have Kaluza-Klein excitations with masses of a few
hundred GeV. We present the constraints on these light fermions from
recent Tevatron searches, and explore interesting discovery channels
at the LHC.

\end{abstract}


\maketitle

\section{Introduction}
\label{Introduction:section}  

Models with warped extra dimensions~\cite{Randall:1999ee} represent a
very exciting alternative to more traditional extensions of the
Standard Model (SM), like supersymmetry.  These models provide not
only a natural solution to the hierarchy problem, but also a very
compelling theory of flavor, when fermions are allowed to propagate in
the bulk~\cite{Grossman:1999ra}.  The large hierarchies among the
different fermion masses arise in a natural way, without inducing new
sources of flavor violation for the first and second generation
fermions.  On the the other hand, large flavor violation effects are
predicted for the third generation fermions, most notably for the top
quark~\cite{Gherghetta:2000qt}.  It has been recently realized that an
enlarged bulk gauge symmetry,
\begin{equation}
SU(2)_L\times SU(2)_R \times U(1)_X~,
\end{equation} 
can act as a custodial symmetry that protects two of the most
constraining observables from large tree-level corrections: the
Peskin-Takeuchi \cite{Peskin:1991sw} $T$
parameter~\cite{Agashe:2003zs} and, if an extra discrete left-right
symmetry $P_{LR}$ is imposed, the anomalous $Z b_L \bar{b}_L$
coupling~\cite{Agashe:2006at}.\footnote{See \cite{Djouadi:2006rk} for
a possible alternative to the custodial protection of the $Z b_L
\bar{b}_L$ coupling.} When the zero modes of the first two generations
are localized far from the IR brane, so as to explain the low-energy
flavor structure, an analysis of the electroweak (EW) precision data
based on the oblique corrections parametrized by $S$ and $T$, together
with the heavy flavor asymmetries and branching ratios, takes into
account the most important effects.  Therefore, at tree level, the
custodial protection of the $T$ parameter and the $Zb_{L}\bar{b}_{L}$
coupling leaves the $S$ parameter as the only relevant constraint.

However, as emphasized in~\cite{Carena:2006bn}, there are calculable
one-loop corrections to the precision electroweak observables, which
can be relevant and should be taken into account.  In fact, for the
choice of quantum numbers that lead to the custodial protection of the
$Zb_{L}\bar{b}_{L}$ coupling, it was shown that the contribution of
these loop corrections yields tight constraints on the parameters of
these models, which in turn can have interesting implications for the
spectrum of Kaluza-Klein (KK) states and their phenomenology.

In Ref.~\cite{Carena:2006bn} we performed an analysis of the
constraints from precision electroweak observables, including one-loop
effects, for a specific model based on gauge-Higgs unification, when
the light fermions are localized near the UV brane.  We obtained a
bound on the mass of the first level KK excitations of the gauge
bosons of about $3$-$4$ TeV, together with light KK quarks with masses
of order of a few hundred GeV, some of them with exotic electric
charges.  Since the $S$ parameter yields one of the most relevant
constraints in these models, one would like to investigate whether
scenarios with the light generation fermions localized near the
conformal point (flat wavefunctions for the zero-modes), where the
couplings of these fermions to the $SU(2)_L$ KK modes are suppressed,
can lead to a better fit~\cite{Cacciapaglia:2004rb}.  Generically,
however, an analysis based on the oblique approximation is not
sufficient in this region, since the couplings of fermions to
$SU(2)_{R}$ gauge bosons tend to induce anomalous, nonuniversal
couplings to the $W$ and $Z$ gauge bosons.  

It is important to emphasize that, due to the custodial symmetry, the
corrections to the $T$ parameter and the $Zb_{L}\bar{b}_{L}$ vertex
cannot receive contributions from higher dimension 5D operators
suppressed by a low cutoff scale and are, therefore, calculable.  In
addition, in any given model these two quantities satisfy a definite
correlation which, in general, may be understood in terms of the
contribution of the lightest KK modes.  The potentially large loop
corrections to the $T$ parameter and the $Zb_{L} \bar{b}_{L}$
coupling, as well as the effects of the associated correlations, must
be considered in any model that makes use of the custodial symmetry.
This includes models of gauge-Higgs unification
\cite{gauge:Higgs:unification,Contino:2003ve,Agashe:2004rs} and models
with a fundamental Higgs or even without a Higgs
\cite{Cacciapaglia:2006gp}.

In this work, we present the results of a global fit to all relevant
EW precision observables, taking into account the correlations among
them as well as possible non-universal effects, in a particular
setting.  We have chosen to test these ideas in the context of
gauge-Higgs unification scenarios, which we find particularly well
motivated theoretically since they address the little hierarchy
problem present in Randall-Sundrum models with a fundamental Higgs.
In addition, this framework naturally leads to light KK fermion
states, often with exotic charges, that makes these scenarios quite
interesting from a phenomenological point of view.

The outline of the paper is as follows.  We introduce the model in
Section~\ref{Model:section} and discuss its main effects on EW
observables in Section~\ref{Effects:section}.  The results of the
global fit are reported in Section~\ref{Fit:section} and we discuss
some simple variations in section~\ref{variations}.  In
Section~\ref{Example:section} we present an interesting example which
allows for light KK fermions for the three generations within the
reach of present and near future colliders.  We discuss the bounds
from EW precision observables in combination with those from direct
searches at the Tevatron.  We also discuss interesting search channels
at the LHC. Finally we conclude in section~\ref{Conclusions:section}.
Some technical results are given in the Appendix.

\section{A Model of Gauge-Higgs Unification}
\label{Model:section}  

Our setup is a 5-dimensional model in a warped background,
\begin{equation}
ds^2=e^{-2ky}\eta_{\mu\nu} dx^\mu  dx^\nu-dy^2~,
\end{equation}
where $0\leq y \leq L$.  The bulk gauge symmetry is $SO(5)\times
U(1)_X$, broken by boundary conditions to $SU(2)_L \times SU(2)_R
\times U(1)_X$ on the IR brane ($y=L$), and to the Standard Model
$SU(2)_L\times U(1)_{Y}$ gauge group on the UV brane
($y=0$)~\cite{Agashe:2004rs}.  The $U(1)_{X}$ charges are adjusted so
as to recover the correct hypercharges, where $Y/2 = T^{3}_{R} +
Q_{X}$ with $T^3_{R}$ the third $SU(2)_{R}$ generator and $Q_{X}$ the
$U(1)_{X}$ charge.  The fifth component of the gauge fields in
$SO(5)/SU(2)_L\times SU(2)_R$ has a (four-dimensional scalar) zero
mode with the quantum numbers of the Higgs boson.  This zero mode has
a non-trivial profile in the extra dimension~\cite{Contino:2003ve},
\begin{equation}
A^{\hat{a}}_5(x,y)=A^{\hat{a}(0)}_5(x) f_H(y) + \ldots~,
\label{A5}
\end{equation}
where $\hat{a}$ labels the generators of $SO(5)/SU(2)_L\times
SU(2)_R$, and is exponentially localized towards the IR brane
\begin{equation}
f_H(y)=\sqrt{\frac{2k}{e^{2k L}-1}} \, e^{2ky}~,
\end{equation}
hence giving a solution to the hierarchy problem.  The dots in
Eq.~(\ref{A5}) stand for massive KK modes that are eaten by the
corresponding KK gauge fields.

The SM fermions are embedded in full representations of the
bulk gauge group.  The presence of the $SU(2)_R$ subgroup of the full
bulk gauge symmetry ensures the custodial protection of the $T$
parameter \cite{Agashe:2003zs}.  In order to have a custodial
protection of the $Zb_L\bar{b}_L$ coupling, the choice
$T^3_R(b_L)=T^3_L(b_L)$ has to be enforced \cite{Agashe:2006at}.  An
economical choice is to let the SM $SU(2)_{L}$ top-bottom doublet
arise from a ${\bf 5}_{2/3}$ of $SO(5)\times U(1)_X$, where the
subscript refers to the $U(1)_X$ charge.  As discussed in
\cite{Carena:2006bn}, putting the SM $SU(2)_{L}$ singlet top in the
same $SO(5)$ multiplet as the doublet, without further mixing, is
disfavored since, for the correct value of the top quark mass, this
leads to a large negative contribution to the $T$ parameter at one
loop.  Hence we let the right-handed top quark arise from a second
${\bf 5}_{2/3}$ of $SO(5)\times U(1)_X$.  The right handed bottom can
come from a ${\bf 10}_{2/3}$ that allows to write the bottom Yukawa
coupling.  For simplicity, and because it allows the generation of the
CKM mixing matrix, we make the same choice for the first two quark
generations.  We therefore introduce in the quark sector three $SO(5)$
multiplets per generation as follows:
\begin{eqnarray}
\begin{array}{c}
\begin{array}{ccccccc}
\xi^{i}_{1L} &\sim&
Q^{i}_{1L} &=& \begin{pmatrix} 
\chi^{u_{i}}_{1L}(-,+) & q^{u_{i}}_L(+,+) \\
\chi^{d_{i}}_{1L}(-,+) & q^{d_{i}}_L(+,+) \end{pmatrix}
&\oplus& u^{\prime i}_L(-,+)~, 
\vspace{3mm}
\\
\xi^{i}_{2R} &\sim&
Q^{i}_{2R} &=& \begin{pmatrix} 
\chi^{u_{i}}_{2R}(+,-) & q^{\prime {u_{i}}}_R(+,-) \\
\chi^{d_{i}}_{2R}(+,-) & q^{\prime {d_{i}}}_R(+,-) \end{pmatrix}
&\oplus& u^{i}_R(+,+)~, 
\end{array}
\vspace{3mm}
\\
\begin{array}{ccccccccccccc}
\xi^{i}_{3R} &\sim&
T^{i}_{1R} &=& \begin{pmatrix}
\psi^{\prime i}_R(-,+) \\
U^{\prime i}_R(-,+) \\
D^{\prime i}_R(-,+) \end{pmatrix}
&\oplus&
T^{i}_{2R} &=& \begin{pmatrix}
\psi^{\prime\prime i}_R(-,+) \\
U^{\prime\prime i}_R(-,+) \\
D^{i}_R(+,+) \end{pmatrix}
&\oplus&
Q^{i}_{3R} &=& \begin{pmatrix} 
\chi^{u_{i}}_{3R}(-,+) & q^{\prime \prime u_{i}}_R(-,+) \\
\chi^{d_{i}}_{3R}(-,+) & q^{\prime \prime d_{i}}_R(-,+) \end{pmatrix},
\end{array}
\end{array}
\label{multiplets}
\end{eqnarray}
where we show the decomposition under $SU(2)_L\times SU(2)_R$.  The
$Q^{i}$'s are bidoublets of $SU(2)_L\times SU(2)_R$, with $SU(2)_L$
acting vertically and $SU(2)_R$ acting horizontally.  The
$T^{i}_{1}$'s and $T^{i}_{2}$'s transform as $({\bf 3}, {\bf 1})$ and
$({\bf 1}, {\bf 3})$ under $SU(2)_L\times SU(2)_R$, respectively,
while $u^{i}$ and $u^{\prime i}$ are $SU(2)_L\times SU(2)_R$ singlets.
The superscripts, $i=1,2,3$, label the three generations.

We also show the boundary conditions on the indicated 4D chirality,
where $-$ stands for Dirichlet boundary conditions.  The $+$ stands
for a linear combination of Neumann and Dirichlet boundary conditions,
that is determined via the fermion bulk equations of motion from the
Dirichlet boundary condition obeyed by the opposite chirality.  In the
absence of mixing among multiplets satisfying different boundary
conditions, the SM fermions arise as the zero-modes of the fields
obeying $(+,+)$ boundary conditions.  The remaining boundary
conditions are chosen so that $SU(2)_{L} \times SU(2)_{R}$ is
preserved on the IR brane, and so that mass mixing terms, necessary to
obtain the SM fermion masses after EW symmetry breaking, can be
written on the IR brane.  It is possible to flip the boundary
conditions on $Q^{i}_{2R}$, consistently with these requirements, and
we will comment on such a possibility in later sections.

As for the leptons, one option is to embed the SM $SU(2)_{L}$ lepton
doublets into the ${\bf 5}_{0}$ representation of $SO(5) \times
U(1)_{X}$ and the $SU(2)_{L}$ charged lepton singlets in a ${\bf
10}_{0}$.  Right-handed neutrinos may come from the $SU(2)_{L} \times
SU(2)_{R}$ singlet in the ${\bf 5}_{0}$, or from a different ${\bf
5}_{0}$, as in the quark sector.  The boundary conditions may then be
chosen in analogy with those in Eq.~(\ref{multiplets}).  A second
possibility is that the leptons, unlike the quarks, arise from the
4-dimensional spinorial representation of $SO(5)$, so that the
SM lepton doublets transform as $({\bf 2}, {\bf 1})$ under
$SU(2)_L\times SU(2)_R$, while the SM lepton singlets
transform as $({\bf 1}, {\bf 2})$.

As remarked above, the zero-mode fermions can acquire EW symmetry
breaking masses through mixing effects.  The most general $SU(2)_L
\times SU(2)_R\times U(1)_X$ invariant mass Lagrangian at the IR brane
--compatible with the boundary conditions-- is, in the quark sector,
\begin{equation}
\mathcal{L}_m = \delta(y-L) \Big[ \bar{u}^\prime_L M_u u_R 
+ \bar{Q}_{1L} M_d Q_{3R} + \bar{Q}_{2L} M_{ud} Q_{3R} + \mathrm{h.c.}
\Big] ~,
\label{localizedmasses}
\end{equation}
where $M_u$, $M_d$ and $M_{ud}$ are dimensionless $3 \times 3$
matrices, and a matrix notation is employed.

\section{Effects on Electroweak Observables}
\label{Effects:section}  

In order to study the effects that the KK excitations of bulk fermions
and gauge bosons have on EW observables, we compute the effective
Lagrangian that results after integrating them out at tree level,
keeping the leading corrections with operators of dimension six.  As
was mentioned in the introduction, some one-loop corrections are also
important and will be included on top of the tree-level effects.  In
fact, in models with custodial protection of the $Zb_{L}\bar{b}_{L}$
coupling, some of the KK fermions become considerably lighter than the
KK gauge bosons, and can give relevant loop level effects as a result
of their strong mixing with the top quark.  The loop contributions to
the EW observables coming from the gauge boson KK excitations are
suppressed due to their larger masses, as well as to the fact that
they couple via the EW gauge couplings, that are smaller than the top
Yukawa coupling.  Thus we expect their one-loop effects on EW
observables to be subleading and we neglect them.

\subsection{Tree-level effective Lagrangian}
\label{EffLag:section}  

In this section, we compute the effective Lagrangian up to dimension
six operators, obtained when the heavy physics in the models discussed
in the previous section is integrated out at tree level.  We will
express the effective Lagrangian in the basis
of~\cite{Buchmuller:1985jz} where the dimension six operators are
still $SU(2)_L \times U(1)_Y$ invariant.  The procedure is the
following.  We integrate out the heavy physics in an explicitly
$SU(2)_L \times U(1)_Y$ invariant way and then use the SM equations of
motion if necessary to write the resulting operators in the basis
of~\cite{Buchmuller:1985jz}.  Only a subset of the 81 operators in
that basis is relevant for EW precision observables, as discussed
in~\cite{Han:2004az}.  At the order we are considering, we can
integrate out independently each type of heavy physics.  The effective
Lagrangian in the basis of~\cite{Buchmuller:1985jz} after integrating
out the heavy gauge bosons reads
\begin{eqnarray}
\Delta \mathcal{L}_{6} &=&
\alpha_h \mathcal{O}_h 
+\alpha^t_{hl} \mathcal{O}^t_{hl} + \alpha^t_{hq}
\mathcal{O}^t_{hq} +
\alpha^s_{hl}\mathcal{O}^s_{hl}
+\alpha^s_{hq}\mathcal{O}^s_{hq}
+
\alpha_{hu}
\mathcal{O}_{hu}
+
\alpha_{hd}\mathcal{O}_{hd}
+
\alpha_{he}\mathcal{O}_{he}
\nonumber \\
&+&
\alpha^t_{ll} 
\mathcal{O}^t_{ll}
+\alpha^t_{lq}  \mathcal{O}^t_{lq}
+\alpha^s_{ll} \mathcal{O}^s_{ll}
+
\alpha^s_{lq} \mathcal{O}^s_{lq} 
+
\alpha_{le} \mathcal{O}_{le} 
+
\alpha_{qe} \mathcal{O}_{qe} 
+
\alpha_{lu} \mathcal{O}_{lu} 
+
\alpha_{ld} \mathcal{O}_{ld} 
\nonumber \\
&+&
\alpha_{ee} 
\mathcal{O}_{ee}
+
\alpha_{eu} \mathcal{O}_{eu} 
+
\alpha_{ed} \mathcal{O}_{ed}+ \ldots ~,
\label{ops}
\end{eqnarray}
where the dots represent other operators that are irrelevant for the
analysis of EW observables.  Here $h$ stands for the SM Higgs, $q$ and
$l$ refer to the $SU(2)_{L}$ doublet quark and leptons, and $u$, $d$,
$e$ refer to the SM $SU(2)_{L}$ quark and lepton singlets.  The list
of the dimension-six operators generated in our model is:
\begin{itemize}
\item Oblique operators
\begin{equation}
\mathcal{O}_h=|h^\dagger D_\mu h|^2~.
\label{TOp}
\end{equation}
\item Two-fermion operators
\begin{eqnarray}
\begin{array}{cclcccl}
\mathcal{O}_{hl}^s &=& 
\mathrm{i} (h^\dagger D_\mu h)
(\bar{l}\gamma^\mu l) + \mathrm{h.c.}~, 
& &
\mathcal{O}_{hl}^t &=& 
\mathrm{i} (h^\dagger \sigma^a D_\mu h)
(\bar{l}\gamma^\mu \sigma^a l) + \mathrm{h.c.}~, \\
\mathcal{O}_{hq}^s &=& 
\mathrm{i} (h^\dagger D_\mu h)
(\bar{q}\gamma^\mu q) + \mathrm{h.c.}~, 
& &
\mathcal{O}_{hq}^t &=& 
\mathrm{i} (h^\dagger \sigma^a D_\mu h)
(\bar{q}\gamma^\mu \sigma^a q) + \mathrm{h.c.}~, \\
\mathcal{O}_{hu} &=& 
\mathrm{i} (h^\dagger D_\mu h)
(\bar{u}\gamma^\mu u) + \mathrm{h.c.}~, 
& &
\mathcal{O}_{hd} &=& 
\mathrm{i} (h^\dagger D_\mu h)
(\bar{d}\gamma^\mu d) + \mathrm{h.c.}~, \\
\mathcal{O}_{he} &=& 
\mathrm{i} (h^\dagger D_\mu h)
(\bar{e}\gamma^\mu e) + \mathrm{h.c.}~.
\end{array}
\label{hpsiOps}
\end{eqnarray}
\item Four-fermion operators
\begin{eqnarray}
\begin{array}{cclcccl}
\mathcal{O}_{ll}^s &=&
\frac{1}{2}(\bar{l}\gamma^\mu l)(\bar{l}\gamma_\mu l)~,
& &
\mathcal{O}_{ll}^t &=&
\frac{1}{2}(\bar{l}\gamma^\mu \sigma^a l)(\bar{l}\gamma_\mu \sigma^a l)~, 
\\
\mathcal{O}_{lq}^s &=&
(\bar{l}\gamma^\mu l)(\bar{q}\gamma_\mu q)~,
& &
\mathcal{O}_{lq}^t &=&
(\bar{l}\gamma^\mu \sigma^a l)(\bar{q}\gamma_\mu \sigma^a q)~, 
\\
\mathcal{O}_{le} &=&
(\bar{l}\gamma^\mu l)(\bar{e}\gamma_\mu e)~,
& &
\mathcal{O}_{qe} &=&
(\bar{q}\gamma^\mu q)(\bar{e}\gamma_\mu e)~, 
\\
\mathcal{O}_{lu} &=&
(\bar{l}\gamma^\mu l)(\bar{u}\gamma_\mu u)~,
& &
\mathcal{O}_{ld} &=&
(\bar{l}\gamma^\mu l)(\bar{d}\gamma_\mu d)~, 
\\
\mathcal{O}_{ee} &=&
\frac{1}{2}(\bar{e}\gamma^\mu e)(\bar{e}\gamma_\mu e)~,
\\
\mathcal{O}_{eu} &=&
(\bar{e}\gamma^\mu e)(\bar{u}\gamma_\mu u)~,
& &
\mathcal{O}_{ed} &=&
(\bar{e}\gamma^\mu e)(\bar{d}\gamma_\mu d)~.
\end{array}
\label{fourFOps}
\end{eqnarray}

\end{itemize}
The coefficients $\alpha_i$ encode the dependence on the different
parameters of our model and their explicit form is given in the
Appendix.

The heavy fermions can be integrated out in a similar
fashion~\cite{delAguila:2000aa}.  However their effects are typically
negligible for all the SM fermions except for the top
quark~\cite{delAguila:2000kb}, whose couplings are irrelevant for the
EW precision observables (except for one-loop
corrections~\cite{DelAguila:2001pu} that will be considered in the
next subsection).  We have nevertheless included all these effects
numerically.~\footnote{There are potentially large tree-level mixing
effects for the bottom quark as well~\cite{Agashe:2004rs}, which
\textit{do} affect the EW precision observable fit.  Such effects are,
however, negligible with the current choice of quantum numbers and
boundary conditions.}

The operator $\mathcal{O}_h$ gives a direct contribution to the $T$
parameter
\begin{equation}
T =-\frac{4\pi v^2}{e^2} \alpha_h
=-\frac{4 \pi v^2}{c^2}[\delta^2_{++}-\delta^2_{-+}] ,\
\label{Ttree}
\end{equation}
where $\alpha_h$ is the coefficient of the corresponding operator as
given in the Appendix, $e$ is the positron charge, $c$ is the cosine
of the weak mixing angle, $v = 174~\rm GeV$ is the Higgs vev, and
$\delta^2_{++}$ and $\delta^2_{-+}$ are functions depending on the
Higgs and KK gauge boson wavefunctions, as defined in
Eq.~(\ref{deltas}).  The (partial) cancellation between
$\delta^2_{++}$ and $\delta^2_{-+}$ in the tree-level contribution to
the $T$ parameter of Eq.~(\ref{Ttree}) is a consequence of the
custodial symmetry.  Also note that the $S$ parameter, generated by
the operator
\begin{equation}
\alpha_{WB} \mathcal{O}_{WB} = \alpha_{WB} \, (h^\dagger \sigma^a h) 
W^a_{\mu\nu} B^{\mu\nu}~, \label{alpha:WB}
\end{equation}
where $S = (32 \pi s c/e^2) v^2 \alpha_{WB}$, is not induced at tree
level in our model.\footnote{Note that this is not in contradiction
with previous claims that a moderate $S$ parameter is generated in
these models.  This contribution to the $S$ parameter comes from a
field redefinition that absorbs a global shift in the gauge couplings
of the light fermions into the oblique $S$ parameter.  Here we do not
do that field redefinition as the shift in the couplings is
automatically included in the global fit.}

\subsection{Large one-loop effects}
\label{Loop:section}  

Although higher dimensional models are nonrenormalizable and many
observables receive contributions from higher-dimension operators
whose coefficients can only be determined by an unspecified UV
completion, it is noteworthy that some of the low-energy observables
are actually insensitive to the UV physics.  This is the case of the
Peskin-Takeuchi $T$ parameter and of the $Zb_{L}\bar{b}_{L}$ coupling
in models with custodial symmetry and the quantum numbers used in this
paper.  In particular, loop contributions to these parameters are
dominated by the KK scale.  This follows simply from the fact that the
assumed symmetries ($SU(2)_{L}\times SU(2)_{R}$ with a discrete
symmetry exchanging $L$ with $R$) and quantum number assignments do
not allow for local 5D counterterms that can contribute to these
observables.  Note, however, that one can write operators that
contribute to the $Zt_{L}\bar{t}_{L}$ coupling.  Although these
symmetries are broken by the boundary conditions at the UV brane, such
breaking is non-local and effectively leads to finite contributions to
the $T$ parameter and the $Zb_{L}\bar{b}_{L}$ coupling at loop
level.\footnote{One can write counterterms that contribute to the $T$
parameter on the UV brane, where the symmetry is reduced to that of
the Standard Model.  However, such effects are suppressed by the
Planck scale, and also by the exponentially small Higgs wavefunction.}

The detailed computation of the leading one-loop contributions to the
$T$ parameter was first performed in~\cite{Carena:2006bn}.  The
important observation made in that work is that the presence of
bidoublets, necessary to protect the tree-level contribution to the
$Zb_{L}\bar{b}_{L}$ coupling, typically induces a negative $T$
parameter at one loop.  There are also contributions from the KK
excitations of the $SU(2)_{L} \times SU(2)_{R}$ singlets that can
alter this result, provided these singlets are relatively light and
mix sufficiently strongly with the top quark.  In this case, a
positive $T$ might be obtained, but also the one-loop contributions to
the $Zb_{L}\bar{b}_{L}$ coupling become sizable and therefore relevant
for the EW fit.

The main one-loop effects, due to heavy vector-like fermions that mix
strongly with the top, can be computed by generalizing the results in
Refs.\cite{Lavoura:1992np,Bamert:1996px}.  We give the detailed
formulas in the Appendix, which can be easily evaluated numerically.
The largest contributions arise from the KK excitations that couple
via the top Yukawa coupling.  In the case of the $T$ parameter, the
quantitative features can be understood from the following types of
contributions:
\begin{itemize}
\item A negative contribution to $T$ from the lightest bidoublet
excitations that violate the custodial symmetry via the boundary
conditions, $Q^3_{1}$ in the notation of Eq.~(\ref{multiplets}).
\item A positive contribution to $T$ from the lightest $SU(2)_{L}
\times SU(2)_{R}$ singlet KK excitations, $u^{\prime 3}$ in the
notation of Eq.~(\ref{multiplets}).
\end{itemize}
In Ref.~\cite{Carena:2006bn} we also gave approximate analytic
formulas for the above contributions.  The expressions for the
bidoublet are somewhat complicated, but the negative contribution
arises from the first KK mode of the
$(\chi^{u_{3}}_{1},\chi^{d_{3}}_{1})$ $SU(2)_{L}$ doublet, that is
lighter and couples more strongly to the Higgs than the lightest KK
mode of the $(q^{u_{3}}_{1},q^{d_{3}}_{1})$ $SU(2)_{L}$ doublet (which
gives a partially compensating positive contribution).  Notice that
the contribution due to the $Q^{3}_{2}$ bidoublet is extremely small,
even when these modes are very light, since the custodial symmetry is
preserved by their boundary conditions.  They can give a nonvanishing
contribution to $T$ only from mixing with other bidoublets that
violate the custodial symmetry.  Our choice for the boundary
conditions of $Q^{3}_{2}$ is motivated by the desire to forbid a
localized mixing mass term between $Q^{3}_{1}$ and $Q^{3}_{2}$, that
would make the $(\chi^{u_{3}}_{1},\chi^{d_{3}}_{1})$ KK modes very
light and their contribution to the $T$ parameter large and negative
(which as we will review below is disfavored by the EW precision data.
See also Ref.~\cite{Carena:2006bn} for further details).  In the
region of parameter space favored by the EW precision data, the
boundary conditions for $Q^{3}_{2}$ result in their KK excitations
easily being in the few hundred GeV range, and present a very
interesting phenomenology (see section~\ref{Example:section}).  It is
in the above sense that we regard very light bidoublets as a rather
well-motivated signature of the scenarios we are studying.

The positive contribution to $T$ from $u^{\prime3}$, mentioned above,
is simply given by
\beqa
\Delta T = T_{\rm top} \frac{2m^{2}_{q^{t}_{0},t}}{M^{2}_{t}}
\left( \ln \frac{M^{2}_{t}}{m^{2}_{\rm top}} - 1 + 
\frac{m^{2}_{q^{t}_{0},t}}{2m^{2}_{\rm top}} \right)~,
\label{Tsinglet}
\eeqa
where $T_{\rm top}$ is the SM contribution from the top quark, with
mass $m_{\rm top}$, $M_{t}$ is the KK mass of $u^{\prime 3}$, and
$m_{q^{t}_{0},t}$ is the EW symmetry breaking mass mixing the lightest
singlet with the SM $(t,b)$ doublet.  There are also terms that arise
from the mixing between the first KK modes of the third generation
$Q^{3}_{1}$ and $u^{\prime 3}$, that can be relevant.

It is important that the dominant fermion loop contributions to the
$Zb_{L}\bar{b}_{L}$ vertex arise from the same set of states as
discussed above.  The contributions coming from the singlet,
$u^{\prime 3}$, are
\begin{equation}
\delta g_{b_L}^s=\frac{\alpha}{16\pi s^2 M_W^2}
\frac{m^4_{q^{t}_{0},t}}{M_t^2} \bigg[1+2 
\frac{m^2_{\rm top}}{m^2_{q^{t}_{0},t}} 
\Big( \ln \left ( \frac{M_t^2}{m^2_{\rm top}}  \right) -1 \Big)
\bigg]~,
\end{equation}
while those coming from $Q^{3}_{1}$ are
\begin{equation}
\delta g_{b_L}^q+\delta g_{b_L}^\chi
=\frac{\alpha}{32\pi s^2 M_W^2}
m_{\rm top}^2 
\bigg[
\frac{m_{q^{t},t}^2}{M_q^2} \ln \left ( \frac{M_q^2}{m_{\rm top}^2}  \right)
-\frac{m_{\chi^{d},t}^2}{M_\chi^2} \ln \left ( \frac{M_\chi^2}{m_{\rm top}^2}  \right)
\bigg]~.
\end{equation}
Here $M_t$, $M_q$ and $M_{\chi}$ are the KK masses of $u^{\prime 3}$,
$q^{u_{3}}_{1}$ and $\chi^{d_{3}}_{1}$, respectively, while
$m_{q^{t},t}$ and $m_{\chi^{d},t}$ are the EW breaking masses that mix
the right-handed top with the lightest KK modes of the two bidoublet
components $q^{u_{3}}_{1}$ and $\chi^{d_{3}}_{1}$, respectively.
Also, $M_{W}$ is the $W$ mass, $\alpha$ is the fine structure constant
and $s$ is the sine of the weak mixing angle.  There are additional
contributions from the mixing between bidoublet and singlet KK modes,
but we do not give the analytic expressions here since they are
somewhat complicated.  The dominant contribution arises from the
singlet, but the mixing terms can also give a relevant effect.

\begin{figure}[t]  
\centerline{\includegraphics[width=0.65\textwidth]{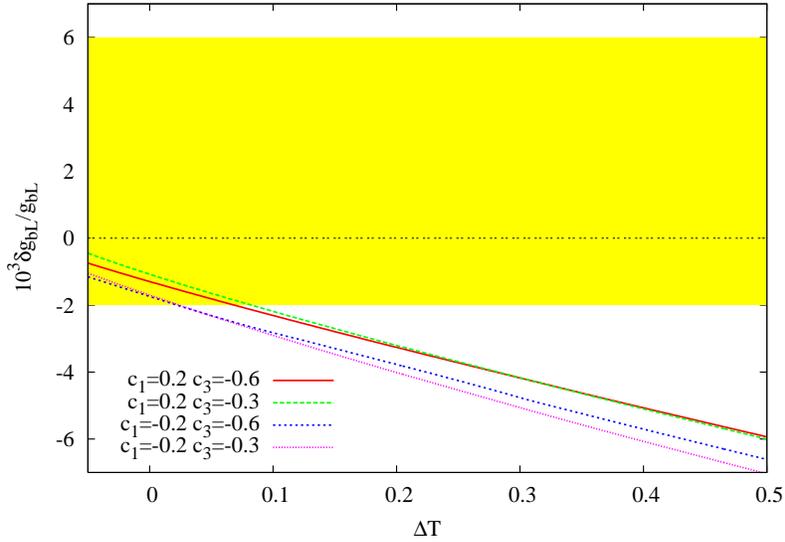}
} \caption{Correlation between the one-loop contributions to the $T$
parameter, denoted by $\Delta T$, and the one-loop contributions to
$\delta g_{b_{L}}/g_{b_{L}}$ in the model of Eq.~(\ref{multiplets}).
We show representative curves for a few values of the left-handed top
quark localization parameter, $c_{1}$, and the bottom quark
localization parameter, $c_{3}$, as the right-handed top localization
parameter, $c_{2}$, is varied.  We take the mass of the first KK
excitation of the $SU(2)_L$ gauge bosons $m_1^\mathrm{gauge} =
3.75~{\rm TeV}$.  The band corresponds to the 2-$\sigma$ bound on
$\delta g_{bL}/g_{bL}$, assuming no large corrections to the $Z b_R
\bar{b}_R$ coupling.}
\label{corrleationTZbb}  
\end{figure}

It should be noted that, although these contributions depend on
several mass and mixing parameters, within the context of an extra
dimensional theory all of these are highly correlated by the shape of
the wavefunctions.  As an example, we show in
Fig.~\ref{corrleationTZbb} the correlation between the one-loop
contributions to $T$ and the $Zb_{L}\bar{b}_{L}$ vertex in the
gauge-Higgs unification scenario based on the $SO(5)\times U(1)_{X}$
gauge symmetry, and with the fermion content given in
Eq.~(\ref{multiplets}).  In particular, we see that in the region
where $T$ becomes positive, the one-loop contribution to the
$Zb_{L}\bar{b}_{L}$ vertex increases, and cannot be neglected in the
EW fit.  In the figure, we did not include the tree-level
contributions to the $T$-parameter from gauge KK mode exchange, which
are subdominant.

Given the importance of these one-loop corrections, we have formally
added them to the effective Lagrangian at the same level as the
tree-level corrections computed in the previous section.\footnote{Note
that the tree-level corrections arise from the gauge sector.  Since
the KK gauge bosons are heavier than the KK fermions, their tree-level
effects can be comparable to the fermion one-loop effects.  We expect
higher-order loop corrections to be subleading, so they can be
neglected.} This is done by simply adding the corresponding results of
Eqs.  (\ref{T:eq}-\ref{Zbb:eq}) to the coefficients of the operators
$\mathcal{O}_h$ and $\mathcal{O}_{hQ}^t$ (here $Q$ represents the
doublet of the third generation).

We have also computed the one-loop contributions to the $S$-parameter.
Although there is no reason to expect the result to be UV insensitive,
one finds that at one loop the sum over KK modes converges fast.  In
the region of parameter space we are interested, the corresponding
contribution to $S$ is not negligible, and we include it as a
contribution to the operator $\mathcal{O}_{WB}$ defined in Eq.
(\ref{alpha:WB}).  We take this as a reasonable estimate of the total
contribution to $S$, but one should keep in mind that, at least in
principle, additional UV contributions to $S$ could have a significant
impact.  Note that, when the light fermions are localized
far from the IR brane, the universal shift in their couplings to the
gauge bosons can be reabsorbed as an additional tree-level
contribution to the $S$ parameter.  Since the tree- and loop-level
contributions to the $S$ parameter have the same sign, it is natural
to assume that there are no particular cancellations when the effects
of the physics above the UV cut-off are included.  In particular, the
$S$ parameter is positive, thus disfavoring the regions of parameter
space that lead to a negative $T$ parameter \cite{Yao:2006px}.

\section{Global fit to Electroweak precision observables}
\label{Fit:section}  

Having computed the leading corrections to the effective Lagrangian in
the model of interest, we can compute the $\chi^2$ function, defined
by
\beqa
\chi^2(\alpha_{i}) &=& \chi^{2}_{\rm min} + (\alpha_{i} - 
\hat{\alpha}_{i}) {\cal M}_{ij}
(\alpha_{j} - \hat{\alpha}_{j})~,
\eeqa
where the $\alpha_{i}$'s, as defined by Eq.~(\ref{ops}), depend on the
fundamental parameters of the model: localization parameters for each
5D fermion multiplet, $c_{\xi_{i}}$, localized fermion mass mixing
parameters, $M_{u}$, $M_{d}$ and $M_{ud}$, as defined in
Eq.~(\ref{localizedmasses}), the gauge couplings, $g_{5}$ and
$g_{5\,X}$, and the overall scale of the new physics, which we take as
$\tilde{k} = k \, e^{-kL}$.  The matrix ${\cal M}_{ij}$ and the vector
$\hat{\alpha}_{i}$ are obtained by performing a global fit to the
experimental data.  We use the fit of Ref.~\cite{Han:2004az}, which
takes into account low-energy measurements, as well as the results
from LEP1, SLD and LEP2.  However, we have not included the NuTeV
results.

Although the model contains a large number of parameters, some of
these, or certain combinations of them, are fixed by the low-energy
gauge couplings, fermion masses and fermion mixing angles.  Also, in
order to avoid dangerous FCNC's we have considered family independent
localization parameters, $c_{\xi_1}^\mathrm{light},
c_{\xi_2}^\mathrm{light}, c_{\xi_3}^\mathrm{light}$, for the
multiplets giving rise to the light SM fermions.  A scan over
parameter space shows that the EW fit favors the light right-handed
(RH) quarks and leptons to be localized near the UV brane
($c_{\xi_2}^\mathrm{light}\sim c_{\xi_3}^\mathrm{light}\sim -0.6$) and
the left-handed (LH) quarks and leptons to be localized close to each
other.  Thus, we will take a common localization parameter for the
light LH quarks and leptons, denoted by $c_\mathrm{light}$, and,
unless otherwise specified, we place the light RH fermions near the UV
brane (we denote their localization parameters by $c_{\rm RH}$).
Although the assumption of family independence is quite important when
the fermions are localized near the IR brane, it is not essential when
the fermions are localized closer to the UV brane (the fermion mass
hierarchies can then be generated by exponential wavefunction
factors).  In particular, if the light fermions are localized close to
the UV brane, the results of our global fit apply even if their
localization parameters are not family universal.  As for the third
quark family, we have allowed independent localization parameters for
the different multiplets: $c_{1}$ for the multiplet giving rise to the
$SU(2)_{L}$ doublet $(t,b)_{L}$, $c_{2}$ for the multiplet giving rise
to $t_{R}$, and $c_{3}$ for the multiplet giving rise to $b_{R}$.

Regarding the localized mass mixing terms of
Eq.~(\ref{localizedmasses}), when the first two generations are
localized near the IR brane, the corresponding terms are extremely
small (of order $m_{f}/\tilde{k}$, where $m_{f}$ is a fermion mass),
and have a negligible effect.  In this case, the only large boundary
mass is $M_u^{33}$, that is fixed by the top quark mass for each value
of $c_1$ and $c_2$.  However, when the light fermions are localized
near the UV brane, the mixing mass terms can be of order one (recall
these are dimensionless parameters).  In this case, they can have an
important effect on the KK spectrum.  Nevertheless, they still have a
negligible effect on the EW fit, for the following reasons.  As
discussed before, there are potentially important contributions from
fermion KK modes both at tree- and loop-level.  The tree-level effects
arise from mixing, after electroweak symmetry breaking, between the
zero-mode and the massive fermion modes, and can affect the couplings
to the gauge bosons of the SM fermions.  Since the region where the
localized masses are of order one corresponds to the case where the
zero-mode fermions are far from the IR brane, and since the mixing
effects are proportional to the overlap between this wavefunction and
the Higgs profile, which is localized near the IR brane, it is easy to
see that the relevant mixing angles are exponentially suppressed.  On
the other hand, when the zero-mode fermions are near the IR brane, the
localized masses are forced to be small due to the smallness of the
light fermion masses, so that the mixing effects are again suppressed
(for the down-type fermions, the custodial symmetry enforces
additional cancellations).  We have checked that these tree-level
effects are always numerically negligible.  The second class of
potentially large effects arises at loop level.  When the zero-mode
fermions are near the IR brane, the loop effects are directly
proportional to the fourth power of the small localized mixing
parameters.  When the zero-mode fermions are localized near the UV
brane, although the loop contributions involving mixing with the zero
mode are exponentially suppressed as above, there are loop
contributions involving only mixing among massive KK states, that are
not necessarily negligible.  However, in this limit the massive KK
spectrum is $SO(5)$ symmetric to a very good approximation.  As a
result, the loop contributions to the $T$ parameter and the
$Zb_{L}\bar{b}_{L}$ vertex discussed in the previous section, due to
the first two generations, are numerically negligible due to the
custodial symmetry.  We have also verified that the $S$ parameter has
only a weak dependence on the localized mixing masses.  Thus, we
conclude that for the purpose of the EW fit analysis, the mixing
masses involving the light generations can be neglected (although, of
course, they are important in reproducing the correct fermion masses
and mixing angles).  Therefore, we are left with six relevant model
parameters: $c_\mathrm{light}$, $c_{\rm RH}$, $c_{1}$, $c_{2}$,
$c_{3}$ and $\tilde{k}$.

It should be noted that in models of gauge-Higgs unification the Higgs
potential --that is induced at loop-level-- is also calculable
\cite{gauge:Higgs:unification}.  Therefore, given the matter and gauge
content of the model, the scale of new physics, $\tilde{k}$, is tied
to the scale of EW symmetry breaking by the gauge and Yukawa couplings
(the latter ones, as determined by the localized mass parameters).
However, it is possible to imagine additional matter content that
could affect the Higgs potential without having an impact on the EW
precision measurements (e.g. 5D fermion multiplets without zero-modes
and with exotic quantum numbers that do not allow mixing with the
SM fermions).  Therefore, we treat $\tilde{k}$ as an
effectively independent parameter.  Given the correlation between
$\tilde{k}$ and the Higgs vev in any such model, one can use our
bounds on $\tilde{k}$ to get an idea of whether the model is excluded
or not (however, if $\tilde{k}$ turns out to be too small, an analysis
that goes beyond the linear treatment of the Higgs couplings used here
might be necessary).  On the other hand, our approach allows us to
apply our bounds to more general models with a bulk Higgs, and where
the Yukawa couplings arise in a similar manner as in gauge-Higgs
unification scenarios.  We will also assume that, as happens in
gauge-Higgs unification scenarios, the Higgs is light, and we have
used a Higgs mass $m_{H} = 120~{\rm GeV}$.

It was shown in Ref.~\cite{Carena:2006bn} that the $T$ parameter in
models with custodial protection of the $Z b_{L} \bar{b}_{L}$ vertex
is negative and non-negligible in a large region of parameter space.
However, it exhibits a strong dependence on the right-handed top
localization parameter, $c_2$, when the right-handed top has a nearly
flat wavefunction, corresponding to $c_2 \sim -0.5$.  In this case,
$T$ can easily reach positive values of order one, so that by
adjusting $c_2$ one can get essentially any value of $T$.  Thus, in
order to reduce the dimension of our parameter space we have chosen to
minimize the $\chi^2$ with respect to $c_2$ for each value of the rest
of the parameters.  Note that this also takes into account the loop
corrections to the $Zb_{L}\bar{b}_{L}$ vertex, since these are
correlated with the $T$ parameter as exemplified in
Fig.~\ref{corrleationTZbb}.  By taking the RH fermions near the UV
brane we are also minimizing with respect to $c_{\rm RH}$.  We have
therefore performed a four-parameter fit and obtained the $2\sigma$
bound on $\tilde{k}$ by varying the $\chi^2$ with respect to the other
three parameters, $c_\mathrm{light}$, $c_1$ and $c_3$.  The first of
these parameters, that determines the localization of the light
fermions, affects directly the tree-level effective Lagrangian
computed in Section~\ref{EffLag:section}, whereas the latter two, that
involve localization of the third quark family, mostly enter the fit
through the one-loop effects discussed in Section~\ref{Loop:section}.

As we mentioned in section~\ref{Model:section}, the SM left-handed
leptons can be embedded either in the vector or spinor representations
of $SO(5)$.  For the first choice, the left-handed SM leptons
transform like $({\bf 2}, {\bf 2})$ under the $SU(2)_{L} \times
SU(2)_{R}$ subgroup, thus allowing for the implementation of the
protection of some of the lepton couplings to the $Z$ gauge boson, as
done in the quark sector.  Such a protection, however, is not as
essential as in the quark sector, since there are no very massive
leptons.  This allows for the second possibility where the SM leptons
transform like $({\bf 2}, {\bf 1})$ or $({\bf 1}, {\bf 2})$ under
$SU(2)_{L} \times SU(2)_{R}$.  As we show below, the bounds are
somewhat relaxed for the second choice.  Thus, we concentrate on this
possibility, and mention the results of the fit when bidoublets are
used for the leptons when appropriate.
  
A scan over parameter space gives a $2\sigma$ lower bound 
\begin{equation}
\tilde{k}\gtrsim 1~\mathrm{TeV}~(95\% ~\mathrm{C.L.})~, 
\end{equation}
which in turns implies a mass for the first gauge KK excitations
$m_1^\mathrm{gauge} \gtrsim 2.5$ TeV. This bound is saturated for
$c_1\approx 0.2-0.3$, $c_{\rm light} \approx 0.48$ and $c_3 \approx -
0.55$ (with the RH light fermions localized near the UV brane and a
nearly flat $t_{R}$ wavefunction with $c_{2} \approx -0.47$).  On the
other hand, when all the light fermions are localized near the UV
brane a bound of $\tilde{k}\gtrsim 1.4~\mathrm{TeV}$ is obtained,
consistent with the result we found in Ref.~\cite{Carena:2006bn} where
a partial fit based on oblique parameters and the $b$ asymmetries and
branching fractions was used.  This confirms the expectation that the
partial fit captures the main effects of the new physics on the EW
precision observables in the case that the light fermions are
localized near the UV brane.

\begin{figure}[t]  
\centerline{\includegraphics[width=0.47\textwidth]{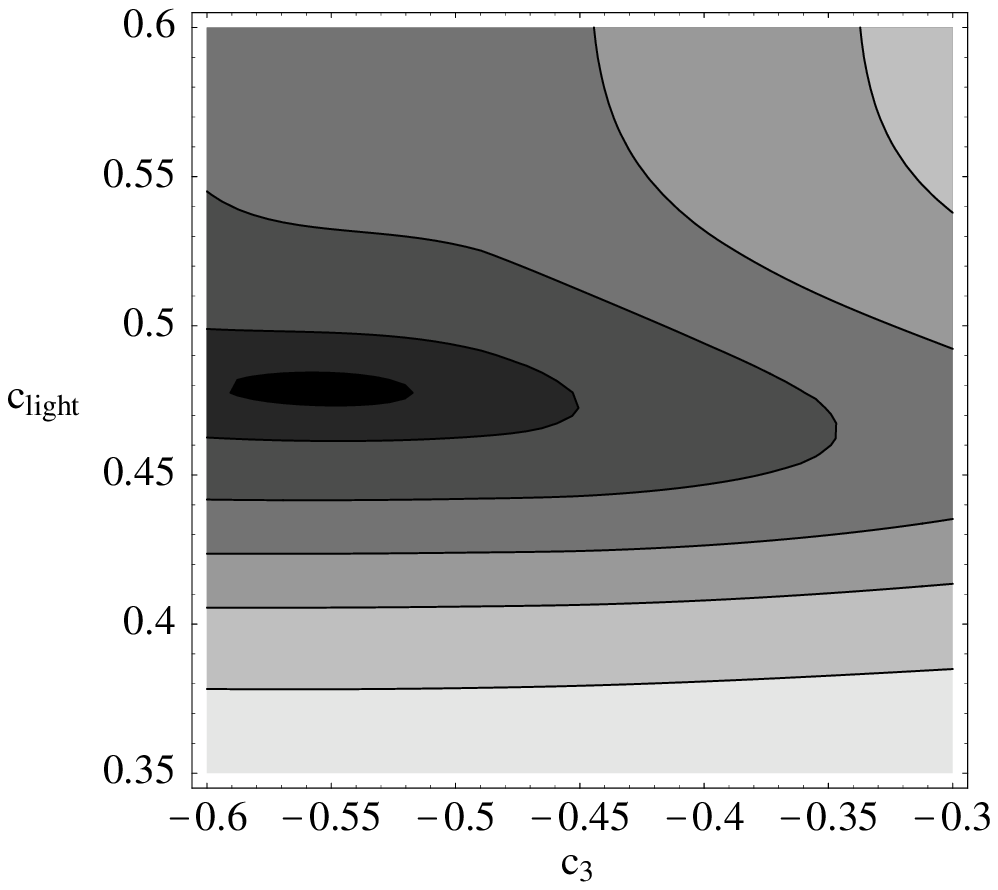}
\includegraphics[width=0.6\textwidth]{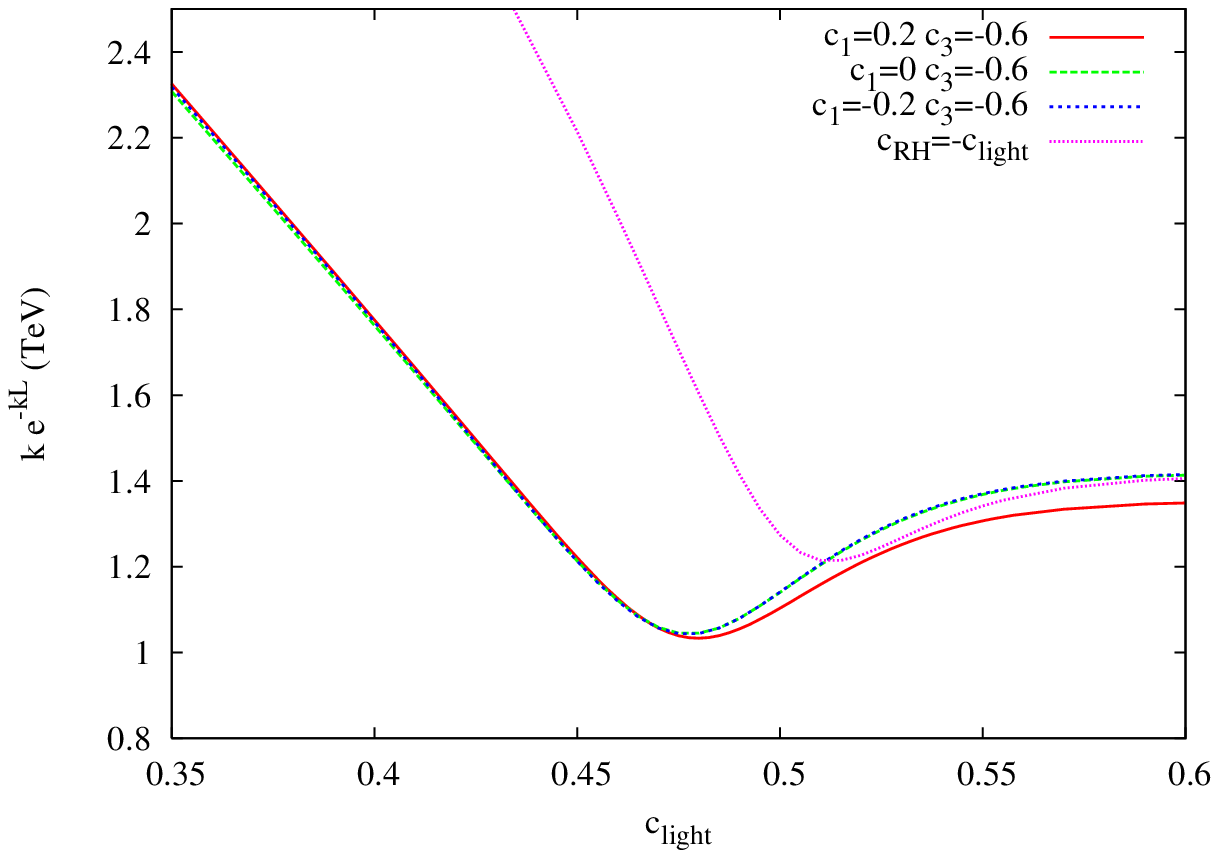}}
\caption{Lower bound on $\tilde{k} = k\,e^{-kL}$ as a function of
$c_3$ and $c_\mathrm{light}$ for fixed $c_1=0.2$ and $c_{\rm RH} =
-0.6$ (left panel).  The different contours, from dark to light,
correspond to $\tilde{k}=1030, 1100, 1300, 1500, 1700$ and $2000 $
GeV, respectively.  The minimum is $\tilde{k}_\mathrm{min}=1$ TeV,
corresponding to $c_3 \approx -0.55$ and $c_\mathrm{light} \approx
0.48$.  In the right panel we show the lower bound on $\tilde{k}$ as a
function of $c_\mathrm{light}$ for fixed $c_{RH} = c_3=-0.6$ and three
values of $c_1$.  We also show the lower bound on $\tilde{k}$ for
$c_{1} = 0.2$ and $c_{3} = -0.6$, assuming $c_{\rm RH} = -c_{\rm
light}$.  The mass of the first gauge KK modes is $m^{\rm gauge}_{1}
\approx 2.5~\tilde{k}$.  }
\label{ktilde:fig}  
\end{figure}

The results are actually quite insensitive to the value of $c_1$, with
slightly better results as we get $Q_1$ farther from the IR brane,
\textit{i.e.} larger $c_1$.  If $Q_1$ is too far from the IR brane,
however, it is not possible to generate the top quark mass, with a
resulting upper bound $c_1\lesssim 0.3$.  In Fig.~\ref{ktilde:fig} we
show, in the left panel, the $2\sigma$ lower bound on $\tilde{k}$ as a
function of $c_3$ and $c_\mathrm{light}$, for fixed $c_1=0.2$, whereas
in the right panel we show the bound on $\tilde{k}$ as a function of
$c_\mathrm{light}$ for fixed $c_3=-0.6$ and three different values of
$c_1=-0.2,0,0.2$, displaying the mild dependence on this latter
parameter.  We also show in the same figure the effect of localizing
the light RH quarks and leptons at the same point as the LH ones.  The
minimum of the fit then shifts to $-c_{\rm RH} =
c_\mathrm{light}\approx 0.51$ with a lower bound $\tilde{k} \gtrsim
1.2$ TeV.

The dependence on the localization of the light fermions is easy to
understand.  The fit is virtually independent of the particular
localization once the conformal point is crossed towards the UV brane,
$c_\mathrm{light}\gtrsim 0.5$, due to the universal couplings of
fermion zero modes to gauge boson KK modes in that case.  There is of
course a limit on how far from the IR brane we can get, given by the
fact that we have to generate the fermion masses.  For instance, the
charm and strange masses force us to take the associated localization
parameters below about 0.6.  This is why we have taken
$c_\mathrm{light} \leq 0.6$ in our plots.  As we have emphasized,
however, the results in that limit are independent of the particular
localization of each light fermion, and we could take the first
generation fermions to be farther away from the IR brane with similar
results.

Also, as is clear from Fig.~2, bringing the light fermions very close
to the IR brane does not improve the fit, due to the strong coupling
to the gauge boson KK modes in that limit.  However, the figure also
shows a minimum when the light fermions are near the conformal point.
It is well known that in this case the (light) fermions decouple from
the KK excitations of the $W$ and $Z$ gauge bosons.  It is
nevertheless important to notice that they do not decouple from the KK
excitations of the $SU(2)_{R}$ gauge bosons and, even near the
conformal point, this leads to non-universal shifts in the gauge
couplings of the SM fermions that cannot be neglected in the fit.  To
illustrate the relevance of such effects, if the custodial protection,
$SU(2)_{V}\times P_{LR}$, is also implemented in the lepton sector,
such non-universal shifts are enough to completely erase the dip in
the $\chi^2$ near the conformal point.  In that case, one finds a
2$\sigma$ lower bound of $\tilde{k}\gtrsim 1.4~\mathrm{TeV}$, obtained
when the light fermions are near the UV brane (this is exactly as in
Fig.  \ref{ktilde:fig}, since in this region the $SU(2)_{R}$ gauge
bosons quickly decouple from the low-energy physics), and the bound
increases monotonically as the light fermions are brought closer to
the IR brane.  Such a feature is a direct result of the fermion
couplings to the $SU(2)_{R}$ gauge bosons as specified by the
embedding into bidoublets of $SU(2)_L \times SU(2)_R$.  We explore
other possibilities in section~\ref{variations}.

Finally, the dependence on the last localization parameter, $c_3$, can
also be easily understood.  In the limit that the light fermions are
near the IR brane ($c_\mathrm{light}\leq 0.5$), the loss of up-down
universality as well as the strong coupling of light fermions to the
gauge boson KK excitations dominate the fit, and therefore the details
of the $b_R$ localization are irrelevant.  This is the reason for the
horizontal contours in the left panel of Fig.~\ref{ktilde:fig} for
$c_\mathrm{light} \lesssim 0.5$.  As the light fermions get farther
from the IR brane, the $b$ asymmetries and branching fractions gain
importance in the fit and therefore there is some dependence on the
value of $c_3$.  The fit shows that the EW precision observables
select the region in which $b_R$ and the RH light fermions are
localized close to the UV brane, whereas the LH light fermions are
near the conformal point slightly towards the IR brane.  In such a
scheme the fermion mass hierarchies can be obtained from the RH
fermion profiles.  Note, however, that the light families, both LH and
RH, could be localized close to the UV brane with only a slightly
tighter bound on $\tilde{k}$.

\section{Effects of simple modifications}
\label{variations}  

The result of the global fit gives an excellent idea of the typical
bounds on the scale of new physics in this class of models.
Nevertheless, they are indirect bounds and they should be interpreted
accordingly.

\begin{figure}[t]  
\centerline{
\includegraphics[width=0.75\textwidth]{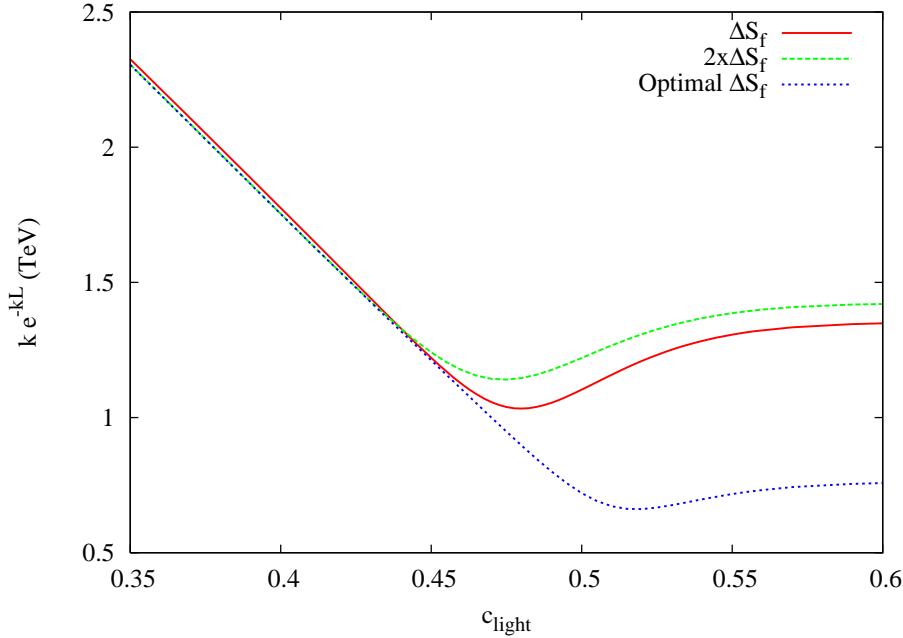}
} \caption{ Lower bound on $\tilde{k} = k\,e^{-kL}$ as a function of
$c_\mathrm{light}$ allowing different contributions to the $S$
parameter in the model of Section~\ref{Fit:section}.  The three lines
correspond to the one-loop contribution from the spectrum in the
model, Eq.~(\ref{S:eq}) (solid line), twice that amount (dashed line)
and a value of $\Delta S_f$ that minimizes the $\chi^2$ for each value
of the parameters (dotted line).  In all cases, $c_{1} = 0.2$ and
$c_{\rm RH} = c_{3} = -0.6$.}
\label{alternatives:fig}  
\end{figure}

In particular, contrary to the $T$ parameter and the
$Zb_{L}\bar{b}_{L}$ coupling, that receive calculable corrections, the
$S$ parameter can get arbitrary corrections from physics at the
ultraviolet cut-off.  This cut-off is warped down to the few TeV scale
and therefore sizable contributions to the $S$ parameter cannot be
ruled out.  To estimate the effect such contribution might have, we
have repeated the global fit with a contribution to the $S$ parameter
that is twice as large as the one we have computed at one loop in our
model.  We have also re-done the fit with an arbitrary contribution to
the $S$ parameter, that we have optimized for each value of the input
parameters.  The results of such fits are shown in
Fig.~\ref{alternatives:fig} with solid line in the case of no extra
contribution to the $S$ parameter beyond the one we have computed,
dashed line for an extra contribution double the one we have computed
in our model, and dotted line in the case that the extra contribution
to the $S$ parameter has been optimized to minimize the $\chi^2$.  The
figure shows that a moderate extra positive contribution to the $S$
parameter worsens the fit slightly whereas optimizing the contribution
leads to a considerably better fit, with a lower bound $\tilde{k}
\gtrsim 650$ GeV (optimal $S$).\footnote{In this case, it is the
observables that depend on the $b$ quark couplings, both at the $Z$
peak and for LEP2, that give all the constraints.} Of course, this
latter possibility is the result of a model tuned to optimize the fit,
most likely requiring a fine-tuned UV completion, and therefore should
not be taken as generic.  Also, for such low values of $\tilde{k}$ the
approximations we have made in linearizing the couplings to the Higgs
in the present gauge-Higgs unification scenarios may have to be
revisited.  Nevertheless, this exercise gives us an idea of how
changes in the model (or like in this case, effects of the UV
completion of our model) can affect these bounds.  In particular, a
negative contribution to the $S$ parameter can be
interesting~\cite{Hirn:2006nt}.
\begin{figure}[ht]  
\centerline{
\includegraphics[width=0.75\textwidth]{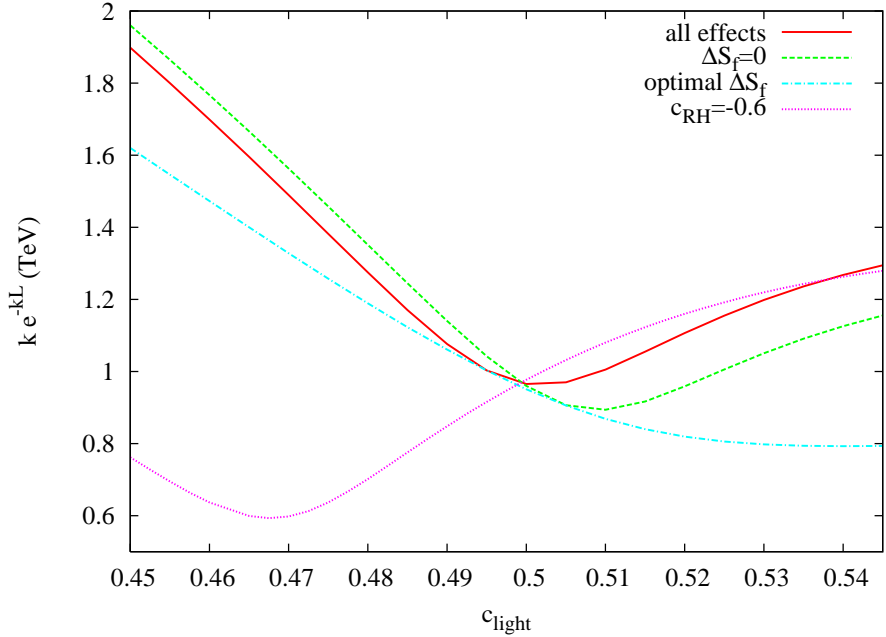}
} \caption{ Lower bound on $\tilde{k} = k\,e^{-kL}$ as a function of
$c_\mathrm{light} = - c_{\rm RH}$ when the light generations arise
from doublets of $SU(2)_{L}$ or $SU(2)_{R}$, including all effects
(solid line), setting the one-loop contributions to $\Delta S_{f}$ to
zero (dashed line), and choosing values of $\Delta S_f$ that minimize
the $\chi^2$ for each value of the parameters (dash-dotted line).  We
also show the lower bound on $\tilde{k}$ when the right-handed light
fermions are localized at $c_{\rm RH} = -0.6$ (near the UV brane), as
a function of the localization parameter for left-handed light
fermions, $c_\mathrm{light}$ (dotted line).  In all cases we have
fixed $c_1=0.2$ and $c_3=-0.6$.  }
\label{alternatives2:fig}  
\end{figure}

A second type of modification is obtained when the quarks of the first
two families --as in the lepton sector-- arise from doublets of
$SU(2)_{L}$ or $SU(2)_{R}$, as opposed to bidoublets of $SU(2)_{L}
\times SU(2)_{R}$.  In this case, it might be difficult to generate
the mixing between the first two quark generations and the third one.
Nevertheless, we have repeated the global fit analysis in such a
scenario, as shown in Fig.~\ref{alternatives2:fig}.  When the LH and
RH fermions have a common localization parameter, $c_{\rm light} =
-c_{\rm RH}$, the fit exhibits a minimum corresponding to a $2\sigma$
bound of $\tilde{k} \approx 1~{\rm TeV}$, again around $c_{\rm light}
\approx 0.5$.  This corresponds to the conformal point, where the
light generations decouple from the KK modes of the $SU(2)_{L}$ gauge
bosons (solid line).  Note, however, that this fit includes the
fermion one-loop contributions to the $S$ parameter, $\Delta S_{f}$,
that are sizable.  If we do not include such loop contributions, the
fit prefers that the light generations be localized somewhat closer to
the UV brane (dashed line).  This is contrary to the naive
expectations, since when the presence of the $SU(2)_{R}$ gauge bosons
is taken into account, the tree-level corrections to the couplings
between the left-handed SM fermions and the gauge bosons vanish at a
point slightly closer to the IR brane.  The result we find can be
explained by observing that, under the assumption $c_{\rm RH} = -
c_{\rm light}$, we cannot simultaneously avoid the corrections to the
couplings involving left- and right-handed fermions, and the global
fit still prefers a region of parameter space where the couplings to
the $SU(2)_{R}$ gauge bosons are somewhat suppressed (fermions closer
to the UV brane).  In fact, when the light RH quarks and leptons are
localized near the UV brane, the fit shows a clear and pronounced
minimum at $c_\mathrm{light} \approx 0.47$.  In that case, one finds a
lower bound on $\tilde{k} \gtrsim 600~{\rm GeV}$, due to an
improvement in $A_{FB}^b$ resulting from a decrease in $A_e$.
Finally, we have re-done the fit, again for
$c_{RH}=-c_\mathrm{light}$, with an arbitrary contribution to the
$S$ parameter, optimized to minimize the $\chi^2$ (dash-dotted line).
As previously discussed, such a scenario could arise from a (possibly
fine-tuned) UV completion.

We therefore conclude that both the calculable loop corrections, as
well as various sources of nonuniversal shifts to the couplings
between fermions and gauge bosons can place important restrictions,
and that a global fit analysis is essential in a broad class of warped
scenarios, whenever the light generations are not close to the UV
brane.  We find that the indirect bounds on $\tilde{k}$ are typically
around a TeV.

\section{Spectrum and phenomenological implications}
\label{Example:section}  

We have seen that a global fit to EW precision observables allows KK
excitations of the SM gauge bosons, together with $W_R^\pm$ and
$Z^\prime$, as light as $m_1^\mathrm{gauge}\sim 2-3$ TeV over a wide
region of parameter space in models with custodial protection of the
$T$ parameter and the $Zb_{L}\bar{b}_{L}$ coupling.  This opens up
exciting possibilities for discovering these particles at the LHC and
measuring their properties~\cite{gaugeKK}.  In particular, the loop
contribution to the $T$ parameter typically singles out a very
specific localization of the third quark family ($t_R$ almost flat and
$(t_L,b_L)$ near the IR brane), that leads to a distinct
phenomenology.~\footnote{This interesting possibility, mentioned for
the first time in \cite{Carena:2006bn}, with the $(t_L,b_L)$ quarks
coupling more strongly to the IR brane than $t_R$, was briefly
discussed in~\cite{gaugeKK}.} The fermionic spectrum is even more
exciting as it can be much lighter than the the spectrum of gauge
boson KK modes.  There are two reasons why KK fermions can be light in
these models.  One is the presence of large brane localized masses,
and the other is the natural appearance of twisted boundary
conditions, $(-,+)$ or $(+,-)$.  Large brane localized masses, needed
to generate the large top mass, are a generic feature of these models.
In principle, one could get the top mass through brane localized
masses that connect either bidoublets or singlets [see
Eq.~(\ref{localizedmasses})].  However, in the case of localized
masses connecting two bidoublets, the light KK bidoublet excitations
will mix strongly with the top quark, inducing a negative $T$
parameter which is disfavored by the EW precision data.  Thus, the top
mass should be predominantly obtained by means of brane localized
masses connecting singlets, and therefore light KK singlets are a
generic prediction in these theories.  On the other hand, light KK
fermion bidoublets can arise from twisted boundary conditions,
provided they do not mix strongly with the zero modes.  In particular,
our boundary conditions for $Q_2$, which ensure no mixing between the
bidoublets $Q_1$ and $Q_2$, give quarks much lighter than $\tilde{k}$
provided that the $t_{R}$ wavefunction is nearly flat (i.e. $c_2 \sim
-0.5$), as required by the EW precision data.

\begin{table}[t]
\begin{tabular}{|c|c|c|c|}
\hline
$q^\prime$ & Q & $\mbox{ }\quad m_{q^\prime}$ (GeV)$\mbox{ }\quad$ & decay  \\ 
\hline
$q_1^1$ & $\frac{2}{3}$ & $\sim 200-500$  & 
$q_1^1 \to Z u, \quad (100\%) $ 
\\
\hline
$q_1^2$ & $\frac{2}{3}$ & $\sim 200-500$  & 
$q_1^2 \to Z c, \quad (100\%) $ 
\\
\hline
$q_2^1$ & $\frac{2}{3}$ & $\sim 200-500$  & 
$q_2^1 \to H u, \quad (100\%) $ 
\\
\hline
$q_2^2$ & $\frac{2}{3}$ & $\sim 200-500$  & 
$q_2^2 \to H c, \quad (100\%) $ 
\\
\hline
\hline
$\chi_2^{u_1}
$ & $\frac{5}{3}$ & $\sim 200-500$  & 
$\chi_2^{u_1} \to W u, \quad (100\%) $ 
\\
\hline
$\chi_2^{u_2}
$ & $\frac{5}{3}$ & $\sim 200-500$  & 
$\mbox{ }\quad\chi_2^{u_2} \to W c, \quad (100\%) \mbox{ }\quad$ 
\\
\hline
\hline
$q^{\prime d_1}$ & $-\frac{1}{3}$ & $\sim 200-500$  & 
$q^{\prime d_1} \to W u, \quad (100\%) $ 
\\
\hline
$q^{\prime d_2}$ & $-\frac{1}{3}$ & $\sim 200-500$  & 
$q^{\prime d_2} \to W c, \quad (100\%) $ 
\\
\hline
\end{tabular}
\caption{\label{light:family:table} 
Electric charges, typical masses and decay channels for the KK
excitations of the first two quark families with masses below 1 TeV.
Here, $q^{i}_{1}$ and $q^{i}_{2}$ are linear combinations of the gauge
eigenstates $q^{\prime u_{i}}$ and $\chi^{d_{i}}_{2}$ of
Eq.~(\ref{multiplets}).}
\end{table}

The typical fermionic spectrum in our model is shown in
Tables~\ref{light:family:table} and \ref{3rd:family:table}.  For each
of the first two families, the four quarks in $Q_2$ of
Eq.~(\ref{multiplets}) have very light KK excitations for
$c_\mathrm{RH}\lesssim -0.5$~\cite{DelAguila:2001pu,Agashe:2004bm}.
As shown in Table~\ref{light:family:table}, there are eight quarks
with almost degenerate masses of a few hundred GeV. Four of them have
charge $2/3$, and two decay almost exclusively to $Z+j$ (where $j$
denotes a jet from an up or charm quark) while the other two decay to
$H+j$.  There are also two light quarks with charge $-1/3$ and two
with exotic charge $5/3$, which decay to $W+j$.

\begin{figure}[t]  
\centerline{\includegraphics[width=0.47\textwidth]{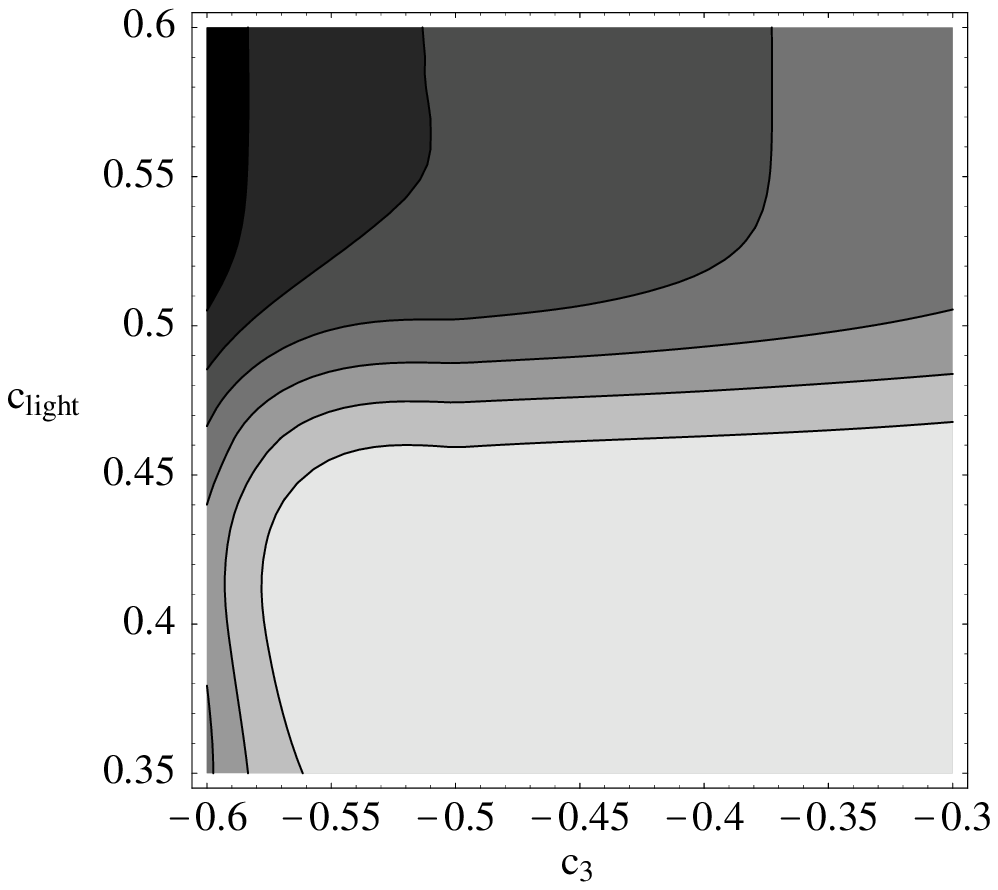}
\includegraphics[width=0.6\textwidth]{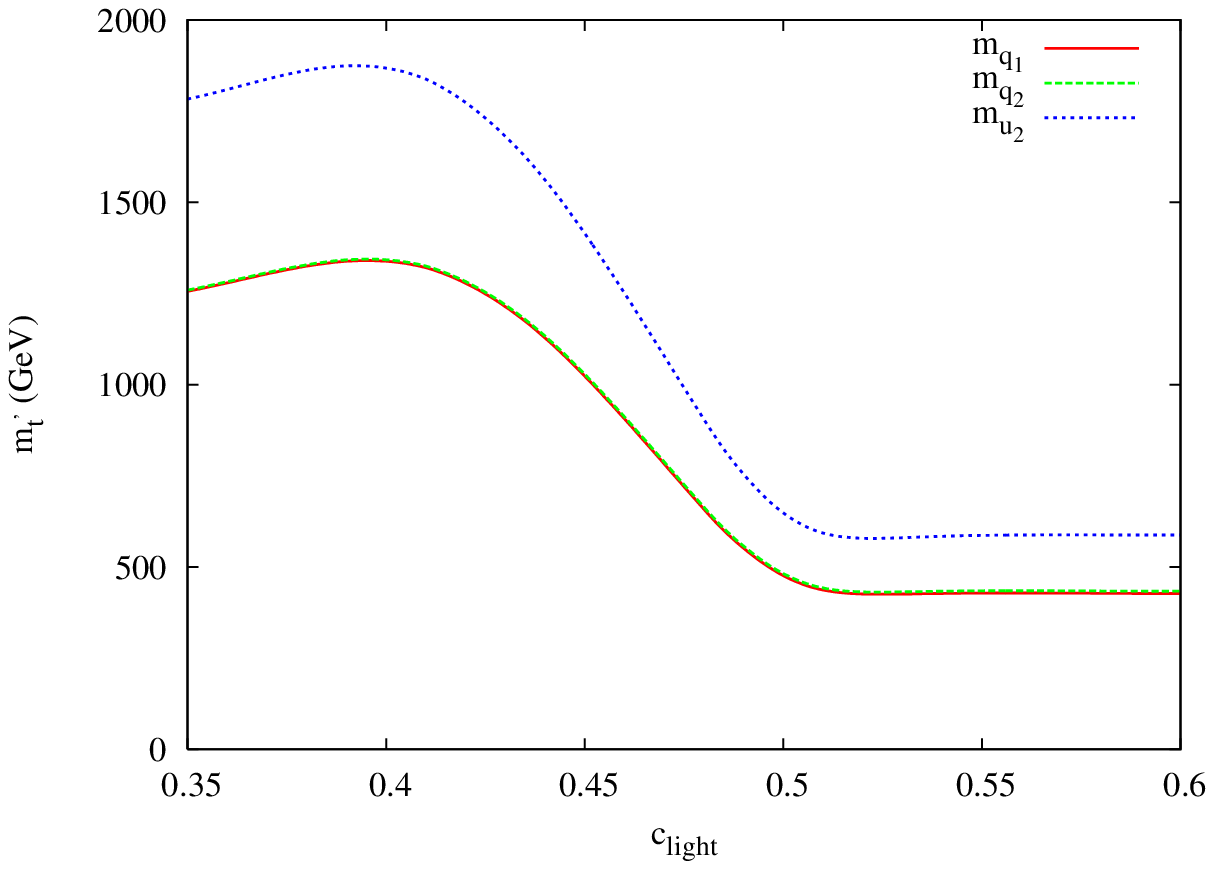}}
\caption{Mass of the first level quarks of the third generation for
the model of Section~\ref{Fit:section} and a $\tilde{k}$ that
saturates the bound, assuming $c_{\rm RH} = - c_{\rm light}$ (dotted
curve in the right panel of Fig.~\ref{ktilde:fig}).  In the left panel
we show the masses of the three degenerate quarks with charges $5/3$,
$2/3$ and $-1/3$ as a function of $c_3$ and $c_\mathrm{light}$, for
fixed $c_1=0.2$.  The different contours, from dark to light,
correspond to $m_1=500, 600, 750, 1000, 1250$ and $1500$ GeV,
respectively.  In the right panel we show the mass of the three
lightest quark KK modes with charge 2/3 as a function of
$c_\mathrm{light}$ for fixed $c_1=0.2$ and $c_3=-0.6$.  }
\label{fermion:spectrum:fig} 
\end{figure}

For the third family, we have three essentially degenerate KK
excitations with charges $5/3$, $2/3$ and $-1/3$.  There is a fourth
KK excitation with charge $+2/3$ and a mass very close to the previous
states.  In Fig.~\ref{fermion:spectrum:fig}~(left), we show the
variation of the masses of the three lightest, degenerate KK quark
excitations --that couple strongly to the third generation-- as
functions of the basic parameters of the model, for fixed $c_1=0.2$
and $\tilde{k}$ saturating the lower bound from the global fit that
assumes a common localization parameter for all the light fermions
(both LH and RH chiralities).  Fig.~\ref{fermion:spectrum:fig}~(right)
shows the masses of the three lightest KK modes with charge $2/3$ as a
function of $c_{\mathrm{light}}$ for fixed values of $c_1 = 0.2$ and
$c_3 = -0.6$.  As seen in the figures, quarks as light as about $400$
GeV are allowed in the region in which the light fermions are near the
conformal point and $b_R$ is near the UV brane.

As an example, we show in Table~\ref{3rd:family:table} the typical
values for the light KK excitations for the third family when
$c_\mathrm{light}=-c_{\rm RH}=0.52$, $c_1=0.2$, $c_2=-0.49$,
$c_3=-0.6$, $\tilde{k}=1.2$ TeV, and localized mass parameters such
that the SM quark masses and CKM matrix are correctly reproduced.
There are three quarks with charge $2/3$, two of them with almost
degenerate masses of about $370$ GeV and the third one with mass of
about $500$ GeV. All of them have decays to $Z+t$, $H+t$ and $W+b$ as
shown in the table.  There are also two other light KK modes, one with
charge $-1/3$ and another with exotic charge $5/3$, with degenerate
masses of order $370$ GeV and which decay almost exclusively to $W+t$.
Note that there is a small but non-negligible probability for the
heavier of the three quarks with charge $2/3$ to decay into the quarks
of mass $\sim 370$ GeV (with either charge).
\begin{table}[t]
\begin{tabular}{|c|c|c|c|}
\hline
$q^\prime$ & Q & $m_{q^\prime}$ (GeV) & decay  \\ 
\hline
$q_1$ & $\frac{2}{3}$ & 369  & 
$\begin{array}{l} 
q_1 \to Z t, \quad (20\%) \\
q_1 \to H t, \quad (60\%) \\
q_1 \to W b, \quad (20\%) 
\end{array}
$ 
\\
\hline
$q_2$ & $\frac{2}{3}$ & 373  & 
$\begin{array}{l} 
q_2 \to Z t, \quad (9\%) \\
q_2 \to H t, \quad (70\%) \\
q_2 \to W b, \quad (21\%) 
\end{array}
$ 
\\
\hline
$u_2$ & $\frac{2}{3}$ & 504  & 
$\begin{array}{l} 
u_2 \to Z t, \quad (13\%) \\
u_2 \to H t, \quad (40\%) \\
u_2 \to W b, \quad (41\%) \\
u_2 \to Z q_1, \quad (1.5\%) \\
u_2 \to W q^{\prime d_3}, \quad (2.5\%)  \\
u_2 \to W \chi^{u_3}_2, \quad (2.\%)  
\end{array}
$ 
\\
\hline
$\chi_2^{u_3}$ & $\frac{5}{3}$ & 369 &
$\chi_2^{u_3}\to W t, \quad (100\%)$
\\
\hline
$q^{\prime d_3}$ & $-\frac{1}{3}$ & 369 &
$q_2^{\prime d_3}\to W t, \quad (100\%)$
\\
\hline
\end{tabular}
\caption{\label{3rd:family:table} 
Electric charges and typical masses and decay channels for the KK
excitations of the third quark family with masses below 1 TeV. We have
fixed $c_{\rm light} = -c_{\rm RH} = 0.52$, $c_1=0.2$, $c_2=-0.49$,
$c_3=-0.6$ and $\tilde{k}=1.2$ TeV. Here, $q_{1}$, $q_{2}$ and $u_{2}$
are mainly admixtures of the gauge eigenstates $q^{\prime u_{3}}$,
$\chi^{d_{3}}_{2}$ and $u^{\prime 3}$ of Eq.~(\ref{multiplets}).}
\end{table}

The rest of the fermion KK modes have masses typically above $1$ TeV.
In the following we shall discuss the potential for searches for the
first level of fermionic excitations at the Tevatron and the LHC.
These models can also have very interesting collider implications in B
and top physics \cite{Agashe:2006wa, Aquino:2006vp} but we postpone
their study for future work.

\subsection{Fermion KK modes at the Tevatron}
\label{lightgenerations}  

\begin{figure}[t]  
\centerline{
\includegraphics[width=0.8\textwidth]{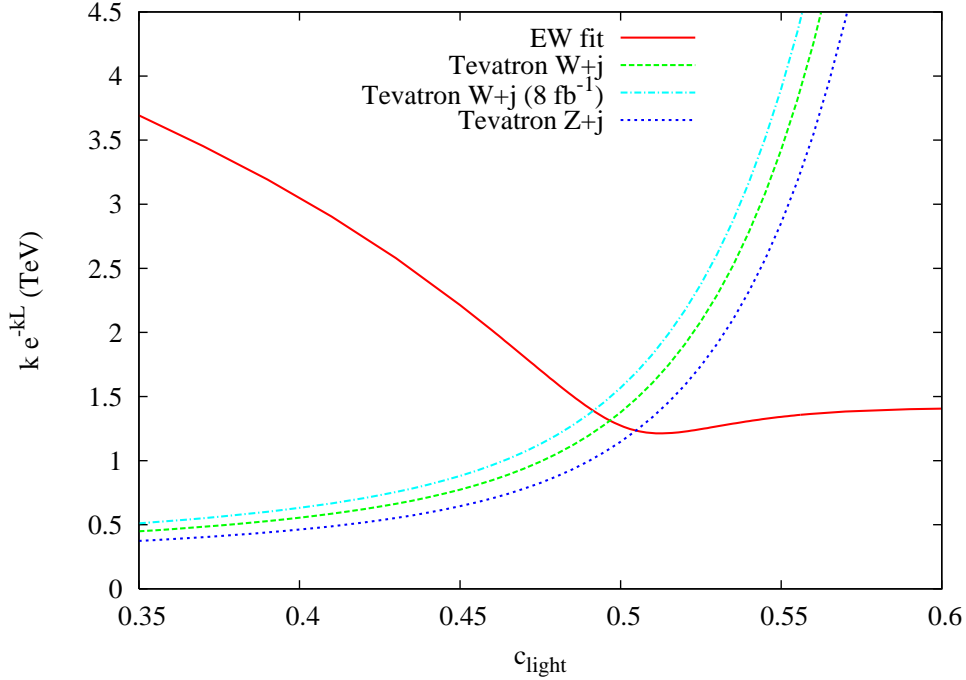}}
\caption{ Lower bound on $\tilde{k} = k\,e^{-kL}$ as a function of
$c_\mathrm{light} = -c_{\rm RH}$ for fixed $c_1=0.2$ and $c_3=-0.6$.
The different lines correspond to the bounds from EW precision
observables (solid), the $W+$ jets analysis at Tevatron with
$0.76~\mathrm{fb}^{-1}$ (dashed), the projected exclusion reach to
$8~\mathrm{fb}^{-1}$ in that same channel (dotted-dashed) and $Z+$
jets analysis at Tevatron (dotted).  The regions below the curves are
excluded.  }
\label{ktildebounds:fig}  
\end{figure}

The Tevatron has excellent capabilities to search for the light KK
excitations of the first two generation quarks shown in
Table~\ref{light:family:table}.  In particular, there are ongoing
Tevatron searches for heavy quarks decaying to $W+j$~\cite{Wj:note}
and $Z+j$~\cite{Zj:note}, which apply directly to our model.  The
first analysis examines the $W+j$ mass spectrum and compares to the
distribution expected from a generic forth-generation top quark.  The
$Z+j$ analysis does not assume any specific model, but rather looks at
the tail of the jet energy distribution for an excess above the SM
expectation.  A similar analysis looks at the $p_T$ distribution of
the $Z$ boson, and in principle could also be sensitive to signals
from our model.  Using the results of these analyses and taking into
account the enhancement factor in the production cross section due to
the multiplicity of quarks (4 in the $W+j$ analysis and 2 in the $Z+j$
one), we obtain the following lower bound on the mass of the light KK
excitations
\begin{equation}
m_q \geq \left \{ \begin{array}{ll} 
325~(410) \mbox{ GeV}, & W+j\mbox{ with }0.76~(\mbox{projected}~8
)~\mathrm{fb}^{-1}~, \\ 
300 \mbox{ GeV}, & Z+j~. \end{array} \right.
\end{equation} 
Figure~\ref{ktildebounds:fig} shows the bound on $\tilde{k}$ from the
fit to EW precision data together with the constraints on our
parameter space that result from these direct searches at the
Tevatron.  The direct search analysis eliminates the region of
parameter space in which the light fermions are localized towards the
UV brane.\footnote{But notice that these direct bounds can be evaded
by switching the boundary conditions for $Q_{2}$ in
Eq.~(\ref{multiplets}) for the first two generations.} When combined
with the EW precision analysis, they slightly strengthen the lower
bound on $\tilde{k}$.

Final states with $Z$ bosons could also lead to a signature with
missing energy and jets.  Searches for squarks and gluinos might be
sensitive to a signal of this type.  It is difficult to relate the
experimental results to our model without performing detailed
simulations.  However, this channel might become interesting with a
sufficiently large integrated luminosity.

We have considered other searches at the Tevatron, such as the
tri-lepton, same-sign di-lepton, and four-lepton searches.  However,
these are usually rather model-dependent and apply cuts which tend to
eliminate our signal.  In particular, leptons reconstructing the $Z$
mass peak as well as jets are typically disallowed.

A very interesting feature is the presence of two light quarks that
decay exclusively into $H+j$.  If these quarks have masses around
$300$ GeV, their production cross section will be of the same order of
magnitude as Higgs production through gluon fusion for a light Higgs.
As there are two such quarks, and two Higgs bosons in every event,
there will be a sizable enhancement to the inclusive Higgs signal.  It
should be noted, however, that some sources of background (such as
$WW+$ jets or $ZZ+$ jets) are also enhanced due to the decays of other
light KK excitations, and a careful analysis of signal and background
is necessary to assess Tevatron prospects for Higgs discovery in this
model.

Finally, the KK excitations of the third quark generation, as shown in
Table~\ref{3rd:family:table}, are on the verge of the projected 
sensitivity for the Tevatron.

\subsection{Fermion KK modes at the LHC}
\label{thirdgeneration}  

The prospects for discovery of the light KK quark excitations of the
first two families are more promising at the LHC. In principle, similar
techniques as those used at the Tevatron could lead directly to a
discovery, although the increase in production comes at the cost of
a larger background from di-boson and top-quark production.

Quarks decaying to $H+j$ enhances Higgs production at the LHC with
respect to the Standard Model.  This cross-section is of the same
order as Higgs gluon fusion production, provided the quarks are not
much heavier than about $400$ GeV. If the Higgs is heavy enough to
decay to $ZZ$ we have the following enhanced contribution to the
inclusive $H\to ZZ$ cross section,
\begin{equation}
\sigma(H\to ZZ)_\mathrm{incl} \approx 2 \sigma(q^\prime
\bar{q}^\prime) 
 B[2-B] + \sigma(gg \to H) B~,
\end{equation}
where we have included the multiplicity of the KK fermions, and where
$B\equiv {\rm BR}(H \to ZZ) \sim 0.02-0.25$ for $m_H \sim 120-200$~GeV
\cite{Carena:2002es}.  For these values of $B$, there can easily be an
enhancement in the inclusive $H\rightarrow ZZ$ of order a few.  Note
that there is also a contribution to the background from quarks
decaying to $Z+j$ although the mass reconstruction in this channel
seems precise enough to efficiently cut that background.  Another
clean channel would be $H \to \gamma \gamma$ for which we can take
advantage of the enhancement of Higgs production without suffering a
larger background from the decays of the lighter KK modes.  Note that
the loop-induced couplings, such as $gg\rightarrow H$ and
$H\rightarrow \gamma\gamma$, are not significantly modified by the
extra quark states.  The reason is that these heavy quark states
receive only a very small contribution to their masses from EW
symmetry breaking, and as a result their effective (diagonal) Yukawa
couplings are very suppressed.

There is an important distinction to be made between the first two
generations and the third one.  While the Tevatron is able to probe
masses on the order of $300$~GeV or higher, these constraints can be
evaded for the first two generations.  If we switch the boundary
conditions of the multiplet $Q_2$ for the first two families, the
excitations of the first two quark generations become heavier, without
affecting the EW fit.  In that case the zero modes for the first two
generations can be localized near the UV brane and the nice features
of the mass generation through wavefunction suppressions preserved,
as well as the flavor universality of the corrections (which become
independent of the particular localization parameter in this limit).
Nonetheless, the KK quark excitations of the third quark generation
remain light, and well within the reach of the LHC
\cite{Aguilar-Saavedra:2005pv,T:littleHiggs} (and possibly of the
Tevatron).

Also for the third generation, the degenerate doublets with different
hypercharges and large mixings with singlets give additional Higgs
production channels which will greatly enhance the signals in
inclusive Higgs searches and searches in the $ttH$ channel.  The Higgs
discovery reach will be even better than the one found in previous
studies \cite{Aguilar-Saavedra:2006gw}, which considered singlets, due
to the enhanced decay ratio to Higgs ($\sim 40 -70 \% $ vs $25 \%$ for
singlets).  

One will notice that nearly all of the final states listed in
Table~\ref{3rd:family:table} result in top quarks in the final state.
This suggests that inclusive top searches would be useful for finding
these particles at the LHC. Another interesting signature might be
multiple jets, some of them with $b$-quarks, and possibly high-$p_T$
leptons or missing energy.

Finally, the exotic quantum numbers of the fermion KK excitations of
the third generation can give rise to spectacular new signatures.  For
instance, the quarks with electric charges $5/3$ and $-1/3$ have
similar decay channels with four $W$'s
\begin{equation}
q^\prime \bar{q}^\prime \to W^+ W^- t \bar{t} \to W^+ W^+ W^- W^- b
\bar{b}~.
\end{equation}
This can lead to a very clean final state $e^+ e^+ \mu^- \mu^- + b
\bar{b} + \not \!\!  E_T$ with very little background.  Furthermore,
for the charge $5/3$, we have two $W$'s of the same charge belonging
to the same decay chain, which could be identified by a pair of
same-sign leptons.  Also, as seen in Table~\ref{3rd:family:table},
$u_{2}$ has a non-negligible branching fraction into the lighter
charge $2/3$ states, which can lead to a spectacular signal with a $6W
b \bar{b}$ final state.

\section{Conclusions}
\label{Conclusions:section}  

Models with warped extra dimensions explain in a compelling way the
hierarchy between the Planck and EW scales.  Bulk fermions provide
also a rationale for the observed hierarchy of fermion masses and the
absence, for the light fermions, of large flavor changing neutral
currents or large new effects from physics above the ultraviolet
cut-off of the theory.  The SM gauge bosons and third generation
quarks, however, have a sizable coupling to heavy states through the
Higgs field.  These couplings induce large corrections to the $T$
parameter and the coupling of the left-handed bottom quark to the $Z$
gauge boson unless some symmetry forbids them.  An enlarged bulk gauge
symmetry can act as a custodial symmetry, $SU(2)_V \times P_{LR}$,
that protects both the $T$ parameter and the $Z b_L \bar{b}_L$
coupling.  When such a symmetry is broken only on the UV brane, these
two observables acquire a distinctive status: they are insensitive to
UV physics except for effects that are suppressed by a scale of order
$M_\mathrm{Pl}$.  This means that, for all practical purposes, they
are calculable.  We find that typically the one-loop corrections to
these observables are sufficiently important that they need to be
included when analyzing the bounds on these models.  Furthermore, for
the fermion quantum numbers required to obtain the SM fermions while
preserving the custodial symmetry, these loop corrections are
correlated.  Thus, they are a generic feature of models with warped
extra dimensions and custodial symmetry $SU(2)_V \times P_{LR}$, no
matter whether the Higgs is a fundamental scalar, the extra
dimensional component of a gauge field (gauge-Higgs unification) or
not present (Higgsless models).  The precise values of the $T$
parameter and the $Zb_{L}\bar{b}_{L}$ coupling are model-dependent,
but we have identified the contributions that are generically present
under the assumption of a custodial symmetry.

We have illustrated these features in a particular model of
gauge-Higgs unification.  We have computed all the relevant tree-level
effects on EW precision observables plus the leading one-loop
corrections to the $T$ parameter and the $Zb_L \bar{b}_L$ coupling.
By performing a global fit to all relevant EW precision observables we
have obtained a lower bound on the masses of the gauge boson KK
excitations of about 2.5 TeV. This bound is saturated when the
left-handed light fermions have nearly flat wavefunctions (the
conformal point), while the right-handed light fermions are localized
near the UV brane.  However, very similar bounds are found in a large
region of parameter space in which all the light fermions are
localized far from the IR brane.  In the latter case, the fit is
dominated by a universal shift of the fermion coupling to the SM gauge
bosons that can be redefined into a pure oblique correction to the $S$
parameter, although the correlation between the $T$ parameter and the
$Zb_{L}\bar{b}_{L}$ coupling also have a noticeable effect.  Contrary
to these two latter observables, the $S$ parameter is not protected by
any symmetry and can receive corrections from physics above the UV
cut-off.  Assuming an optimal correction to the $S$ parameter from UV
physics, such that the $\chi^2$ of the fit is minimized with respect
to $S$ for each point in parameter space, we have obtained a lower
bound on the mass of the gauge boson KK modes of $\sim 1.6$ TeV, which
is completely dominated by the observables in the $b$ sector and is
therefore difficult to evade (as those observables are dominated by
the $Z b_L \bar{b}_L$ coupling that is calculable in these models).

Regarding the fermionic spectrum, there can be a wealth of new
vector-like quarks with exotic quantum numbers and masses as low as a
few hundred GeV. These modes can be light enough for the Tevatron to
have started probing part of the parameter space.  We have discussed
the bounds that current Tevatron analyses place on our model.
Interestingly enough, some of these modes have exotic decay channels,
for instance some of them decaying essentially $100 \%$ into $H$ plus
jets.  This opens up interesting prospects for Higgs physics both at
the Tevatron and the LHC. Heavy quarks decaying to $W+j$ or $Z+j$ have
been searched for at the Tevatron, with current limits of about $325$
GeV and $300$ GeV, respectively, for the quark multiplicities present
in our model.  Excitations of the third generation quarks can have
masses of order $400$ GeV that might be within reach of the Tevatron.
They typically decay to third generation quarks with non-standard
branching ratios, naturally enhancing Higgs production.  Decays to top
quarks through gauge bosons induce a very interesting decay chain with
four gauge bosons ($4W$ or $2W+2Z$) and two $b$'s, as well as a
possible final state with six $W$'s and two $b$'s, that would give a
spectacular signal at the LHC. In particular, heavy quarks (with
typical masses of order $500$ GeV) with electric charge $5/3$ produce
two same sign $W$ in each decay chain whereas those with charge $-1/3$
will give one $W$ of each sign per chain.  The process
\begin{equation}
pp \to q_{5/3,-1/3} \bar{q}_{5/3,-1/3} \to W^+ t W^- \bar{t} 
\to W^+ W^+ b W^- W^- \bar{b} \to \mu^+ \mu^+ e^- e^- b \bar{b} \not
\! \!E_T,
\end{equation}
would lead to an easy discovery of these modes with almost no
background.

\textbf{Note added:} During the final stages of this article, we
received references \cite{Cacciapaglia:2006mz} and
\cite{Contino:2006qr}, which partially overlap with ours in the study
of different aspects of models with warped extra dimensions and
custodial protection of the $Zb_L \bar{b}_L$ coupling.  However they
do not discuss the importance of the calculable one-loop corrections
to the $T$ parameter or the $Z b_L \bar{b}_L$ coupling, nor perform a
detailed global fit to the EW precision observables.

\begin{acknowledgments}
We would like to thank Z.~Han and W.~Skiba for useful comments
on the fits in \cite{Han:2004az}.  We would also like to thank
J.~Conway, R.~Erbacher and M.~Schmitt for very useful discussions
regarding the Tevatron searches and projections.  C.E.M. Wagner thanks
N. Shah and A. Medina for useful comments.  Work at ANL is supported
in part by the US DOE, Div.\ of HEP, Contract No.  DE-AC-02-06CH11357.
Fermilab is operated by Fermi Research Alliance, LLC under Contract
No.  DE-AC02-07CH11359 with the United States Department of Energy.
E.P. was supported by DOE under grant No.  DE-FG02-92ER-40699.
\end{acknowledgments}

\appendix*

\section{4D Effective theory}

In this appendix we give the details of the matching between the 5D
theory and the 4D theory used to perform the global fit analysis.  We
compute the dimension-6 operators and put them in the form of
Eq.~(\ref{ops}).

\subsection{Integration of heavy gauge bosons}

We can perform the tree-level integration of heavy gauge bosons in an
$SU(2)_L \times U(1)_Y$ gauge invariant way by splitting the full
covariant derivative into a \textit{Standard Model} part and a part
involving heavy physics,
\begin{equation}
D_\mu^\mathrm{full}=D_\mu-\mathrm{i} 
\Big[ g_{5\,L} 
\tilde{W}^a_{L\mu} T^a_L +g_{5\,R} \tilde{W}^b_{R\mu}
  T^b_R
+g_5^\prime Y \tilde{B}_\mu + g_{5\,Z^\prime} Q_{Z^\prime}
\tilde{Z}^\prime_\mu \Big]~,
\end{equation}
where $D_\mu$ represents the SM covariant derivative and we use tildes
to denote the massive KK components of the 5D fields.  In the above,
$a=1,2,3$ label the $SU(2)_{L}$ gauge bosons, $b=1,2$ label the
charged $SU(2)_{R}$ gauge bosons, and $B_\mu$ and $Z^\prime_\mu$ are
the following two combinations of neutral gauge bosons
\bea
B_\mu=\frac{g_{5\,X} W^3_{R\, \mu}
+g_{5\,R} X_\mu}{\sqrt{g^2_{5\,R}+g^2_{5\,X}}}~, \quad \quad \quad
Z^\prime_\mu=\frac{g_{5\,R} W^3_{R\, \mu}
-g_{5\,X} X_\mu}{\sqrt{g^2_{5\,R}+g^2_{5\,X}}}~, 
\eea
with $g_{5\,R}, g_{5\,X}$ the five-dimensional coupling constants of
the $SU(2)_{R}$ and $U(1)_{X}$ groups,\footnote{In models that
incorporate the $P_{LR}$ symmetry, including the gauge-Higgs
unification model based on $SO(5)\times U(1)_{X}$ studied in the main
text, one has $g_{5\,R} = g_{5\,L}$.} respectively, while the
hypercharge and $Z^\prime$ gauge couplings are
\beq
g_5^\prime=\frac{g_{5\,R}\,g_{5\,X}}{\sqrt{g^2_{5\,R}+g^2_{5\,X}}}~,
\quad \quad \quad
g_{5\,Z^\prime}=\sqrt{g^2_{5\,R}+g^2_{5\,X}}~.
\eeq 
The charges are
\begin{equation}
\frac{Y}{2}=T^3_R+Q_X~, \quad \quad \quad Q_{Z'}=\frac{g_{5\,R}^2 
T^3_R - g_{5\,X}^2 Q_X}
{g_{5\,R}^2+g_{5\,X}^2}~,
\label{YQZ}
\end{equation} 
so that the electric charge is
\beqa
Q =T^3_L+ T^{3}_{R} + Q_X~.
\label{charge}
\eeqa

The Lagrangian involving heavy fields then reads (terms with two heavy
fields except for kinetic terms give higher order corrections and are
therefore not written)
\begin{eqnarray}
\Delta \mathcal{L} &=&
\frac{1}{2} \tilde{W}^a_{L\mu} \mathcal{O}^{\mu\nu} \tilde{W}^a_{L\nu}
+\frac{1}{2} \tilde{W}^b_{R\mu} \mathcal{O}^{\mu\nu} \tilde{W}^b_{R\nu}
+\frac{1}{2} \tilde{B}_{\mu} \mathcal{O}^{\mu\nu} \tilde{B}_{\nu}
+\frac{1}{2} \tilde{Z}^\prime_{\mu} \mathcal{O}^{\mu\nu} \tilde{Z}^\prime_{\nu}
\nonumber \\
&+&
g_{5\,L} \tilde{J}^{a\mu}_L \tilde{W}^a_{L\mu}
+g_{5\,R} \tilde{J}^{b\mu}_R \tilde{W}^b_{R\mu}
+g^{\prime}_5\tilde{J}^{\mu}_Y \tilde{B}_{\mu}
+g_{5\,Z^\prime}\tilde{J}^{\mu}_{Z^\prime} \tilde{Z}^\prime_{\mu}~,
\end{eqnarray}
where 
\beq
\label{gaugeoperator}
{\cal{O}}^{\mu\nu} \equiv \left[ \eta^{\mu \nu} \partial^2 - 
\partial^\mu \partial^\nu +
 \eta^{\mu\nu} \partial_y \left( e^{-2ky} \partial_y \right) \right]~,
\eeq
and the effective currents read
\begin{eqnarray}
\tilde{J}^{a\mu}_L & =& 
e^{-2 \sigma}
[(T^a_L h)^\dagger \mathrm{i}D^\mu h + \mathrm{h.c.}]
+e^{-3 \sigma}\bar{\psi} \gamma^\mu T^a_L \psi~,
\\
\tilde{J}^{b\mu}_R & =& 
e^{-2 \sigma}
[(T^b_R h)^\dagger \mathrm{i}D^\mu h + \mathrm{h.c.}]
+e^{-3 \sigma}\bar{\psi} \gamma^\mu T^b_R \psi~,
\\
\tilde{J}^{\mu}_Y & =& 
e^{-2 \sigma}
[(Y h)^\dagger \mathrm{i}D^\mu h + \mathrm{h.c.}]
+e^{-3 \sigma}\bar{\psi} \gamma^\mu Y \psi~,
\\
\tilde{J}^{\mu}_{Z^\prime} & =& 
e^{-2 \sigma}
[(Q_{Z^\prime} h)^\dagger \mathrm{i}D^\mu h + \mathrm{h.c.}]
+e^{-3 \sigma}\bar{\psi} \gamma^\mu Q_{Z^\prime}  \psi~.
\end{eqnarray}
The equations of motion for the heavy fields can now be easily written
and solved. For instance, for $\tilde{W}^a_L$, the equations of motion
are
\begin{equation}
\mathcal{O}^{\mu\nu} \tilde{W}^a_{L\nu} = - g_{5\,L} \tilde{J}^{a\mu}_L~,
\label{diffeqn}
\end{equation}
with solution 
\begin{equation}
\tilde{W}^a_{L\mu}(p;y) =g_{5\,L} \int_0^L \mathrm{d}y'\,
\widetilde{G}^{(++)}_{\mu\nu}(p;y,y') \tilde{J}^{a\nu}_L(p;y')~,
\end{equation}
where $\widetilde{G}^{(++)}_{\mu\nu}$ is the propagator for the KK
modes obeying $(+,+)$ boundary conditions (the inverse of the
differential operator in Eq.~(\ref{diffeqn}) with the zero-mode
subtracted), and $p$ is the 4-dimensional momentum.  Inserting the
solution for all the heavy modes back into the Lagrangian we obtain
the following dimension six effective Lagrangian
\begin{eqnarray}
\Delta \mathcal{L}_6&=&
\frac{1}{2} \int_0^L \mathrm{d}y \mathrm{d}y^\prime 
\, \Big[ 
g^2_{5\,_L} 
\tilde{J}^{a\mu}_L \widetilde{G}^{(++)}_{\mu\nu} \tilde{J}^{a\nu}_L
+
g^{\prime\,2}_5 
\tilde{J}^{\mu}_Y \widetilde{G}^{(++)}_{\mu\nu} \tilde{J}^{\nu}_Y
\nonumber \\
&&
\phantom{\frac{1}{2} \int_0^L \mathrm{d}y \mathrm{d}y^\prime}
+g^2_{5\,_R} 
\tilde{J}^{b\mu}_R \widetilde{G}^{(-+)}_{\mu\nu} \tilde{J}^{b\nu}_R
+g^2_{5\,Z^\prime} 
\tilde{J}^{\mu}_{Z^\prime} \widetilde{G}^{(-+)}_{\mu\nu}
\tilde{J}^{\nu}_{Z^\prime} 
\Big]~. 
\label{eff:lag:before}
\end{eqnarray}
Note that these are already operators of dimension six. Thus, the
propagators have to be evaluated at zero momentum. The relevant
expression is
\begin{equation}
\widetilde{G}_{\mu\nu}(p=0,y,y')=\eta_{\mu\nu} \widetilde{G}_{p=0}(y,y') +
\mathcal{O}(p^2),
\end{equation}
where for $(+,+)$ boundary conditions,
\beqa
\widetilde{G}^{(++)}_{p=0}(y,y') &=& \frac{1}{4k(kL)} \bigg\{ \frac{1-e^{2k L}}{kL} +
e^{2ky_{<}} (1-2ky_{<}) + e^{2ky_{>}} \left[ 1+ 2k (L-y_{>})\right]
\bigg\}~,
\label{Gpp}
\eeqa
while for $(-,+)$ boundary conditions
\beqa
\widetilde{G}^{(-+)}_{p=0}(y,y') &=& - \frac{1}{2k} 
\left[ e^{2ky_{<}} - 1 \right]~.
\label{Gmp}
\eeqa
Here $y_{<}$ ($y_{>}$) denote the smallest (largest) of $y$ and $y'$,
the fifth-dimensional coordinate.

The full $y$ dependence of the effective Lagrangian can be encoded in
the following coefficients,
\begin{eqnarray}
\delta_{++}^2 & = &
\frac{L}{2} \int_0^L \mathrm{d}y\mathrm{d}y^\prime \, 
e^{-2 k y} f_H^2(y)
\widetilde{G}^{(++)}_0(y,y^\prime)
e^{-2 k y^\prime} f_H^2(y^\prime), 
\label{deltas}
\\
G^{++}_\psi & = &
\frac{1}{2} \int_0^L \mathrm{d}y\mathrm{d}y^\prime \, 
|f_\psi(y)|^2
\widetilde{G}^{(++)}_0(y,y^\prime)
e^{-2 k y^\prime} f_H^2(y^\prime), \\
G^{++}_{\psi\tilde{\psi}} & = &
\frac{1}{L} \int_0^L \mathrm{d}y\mathrm{d}y^\prime \, 
|f_\psi(y)|^2
\widetilde{G}^{(++)}_0(y,y^\prime)
|f_{\tilde{\psi}}(y^\prime)|^2,
\end{eqnarray}
with similar definitions for $\delta_{-+}^2$, $G^{-+}_\psi$ and
$G^{-+}_{\psi\tilde{\psi}}$ in terms of the propagator of
Eq.~(\ref{Gmp}).  We have used the $y$-dependence of the fermion and
Higgs zero modes
\begin{eqnarray}
\psi(x,y)&=&\frac{e^{3\sigma/2}}{\sqrt{L}} f_\psi(y) \psi(x) + \ldots,
\\
h(x,y)&=&f_H(y) h(x) + \ldots~,
\end{eqnarray}
with the Higgs field, $h(x)$, written here as a doublet of
$SU(2)_{L}$.  Technically, the fermionic dependence is more
complicated due to the non-trivial mass mixing on the brane.  The
analysis of Ref.~\cite{Han:2004az}, however, assumes flavor
universality for the first two families and that will actually be a
very good approximation for the range of parameters we will consider
in the global fit (otherwise large flavor violation involving the
first two families would be generated, in gross conflict with
experimental data).

Note that, even after evaluation of the propagators at zero momentum
and integration over the extra dimension, the effective Lagrangian in
Eq.(\ref{eff:lag:before}) is not yet in the basis
of~\cite{Buchmuller:1985jz}.  Simple manipulations of the operators
involving integration by parts and use of the completeness of the
Pauli matrices takes us to the desired basis.  The resulting effective
Lagrangian then reads,
\begin{eqnarray}
\Delta \mathcal{L}_{6} &=&
\alpha_h \mathcal{O}_h 
+\alpha^t_{hl} \mathcal{O}^t_{hl} + \alpha^t_{hq}
\mathcal{O}^t_{hq} +
\alpha^s_{hl}\mathcal{O}^s_{hl}
+\alpha^s_{hq}\mathcal{O}^s_{hq}
+
\alpha_{hu}
\mathcal{O}_{hu}
+
\alpha_{hd}\mathcal{O}_{hd}
+
\alpha_{he}\mathcal{O}_{he}
\nonumber \\
&+&
\alpha^t_{ll} 
\mathcal{O}^t_{ll}
+\alpha^t_{lq}  \mathcal{O}^t_{lq}
+\alpha^s_{ll} \mathcal{O}^s_{ll}
+
\alpha^s_{lq} \mathcal{O}^s_{lq} 
+
\alpha_{le} \mathcal{O}_{le} 
+
\alpha_{qe} \mathcal{O}_{qe} 
+
\alpha_{lu} \mathcal{O}_{lu} 
+
\alpha_{ld} \mathcal{O}_{ld} 
\nonumber \\
&+&
\alpha_{ee} 
\mathcal{O}_{ee}
+
\alpha_{eu} \mathcal{O}_{eu} 
+
\alpha_{ed} \mathcal{O}_{ed}+ \ldots ~,
\end{eqnarray}
where the list of operators was given in
Eqs.~(\ref{TOp})-(\ref{fourFOps}), and the different coefficients have
the following expressions:
\beq
\begin{array}{cclcccl}
\alpha_h&=&g^{\prime\, 2} [\delta^2_{++} -\delta^2_{-+}]~, 
\\ 
\alpha^t_{hl}&=&
\frac{g_L^2}{2} G_l^{++},\quad
& & \alpha^t_{hq}&=&
\frac{g_L^2}{2} G_q^{++}~,
\\
\alpha^s_{hl}&=&-\frac{g^{\prime \,2}}{2}G^{++}_l
+g_R^2 Q_{Z^\prime}(l) G^{-+}_l~,
\quad
& & \alpha^s_{hq}&=& 
\frac{g^{\prime \,2}}{6}G^{++}_q
+g_R^2 Q_{Z^\prime}(q) G^{-+}_q~, 
\\
\alpha_{hu}&=&\frac{2g^{\prime \,2}}{3}G^{++}_u
+g_R^2 Q_{Z^\prime}(u) G^{-+}_u~,
\quad
& & \alpha_{hd}&=&-\frac{g^{\prime \,2}}{3}G^{++}_d
+g_R^2 Q_{Z^\prime}(d) G^{-+}_d~,
\\
\alpha_{he}&=&-g^{\prime \,2}G^{++}_e
+g_R^2 Q_{Z^\prime}(e) G^{-+}_e~, 
\\
\alpha^t_{ll}&=&g_L^2 G_{ll}^{++}~,
\quad 
& & \alpha^t_{lq}&=&g_L^2 G_{lq}^{++}~,
\\
\alpha^s_{ll}&=&\frac{g^{\prime\,2}}{4} G_{ll}^{++}
+g^2_{Z^\prime} Q_{Z^\prime}^2(l)G_{ll}^{-+}~,
\quad
& & \alpha^s_{lq}&=&-\frac{g^{\prime\,2}}{12} G_{lq}^{++}
+g^2_{Z^\prime} Q_{Z^\prime}(l) Q_{Z^\prime}(q)G_{lq}^{-+}~,
\\
\alpha_{le}&=&\frac{g^{\prime\,2}}{2} G_{le}^{++}
+g^2_{Z^\prime} Q_{Z^\prime}(l) Q_{Z^\prime}(e)G_{le}^{-+}~,
\quad
& & \alpha_{qe}&=&-\frac{g^{\prime\,2}}{6} G_{qe}^{++}
+g^2_{Z^\prime} Q_{Z^\prime}(q) Q_{Z^\prime}(e)G_{qe}^{-+}~,
\\
\alpha_{lu}&=&-\frac{g^{\prime\,2}}{3} G_{lu}^{++}
+g^2_{Z^\prime} Q_{Z^\prime}(l) Q_{Z^\prime}(u)G_{lu}^{-+}~,
\quad
& & \alpha_{ld}&=&\frac{g^{\prime\,2}}{6} G_{ld}^{++}
+g^2_{Z^\prime} Q_{Z^\prime}(l) Q_{Z^\prime}(d)G_{ld}^{-+}~,
\\
\alpha_{ee}&=&g^{\prime\,2} G_{ee}^{++}
+g^2_{Z^\prime} Q_{Z^\prime}(e)^2 G_{ee}^{-+}~,
\quad
& & \alpha_{eu}&=&-\frac{2g^{\prime\,2}}{3} G_{eu}^{++}
+g^2_{Z^\prime} Q_{Z^\prime}(e) Q_{Z^\prime}(u)G_{eu}^{-+}~,
\\
\alpha_{ed}&=&\frac{g^{\prime\,2}}{3} G_{ed}^{++}
+g^2_{Z^\prime} Q_{Z^\prime}(e) Q_{Z^\prime}(d)G_{ed}^{-+}~,
\end{array}
\eeq
with $g_{L} = g_{5L}/\sqrt{L}$, and similarly for the other gauge
couplings.  Here we use the notation $Q_{Z'}(\psi)$ to denote the
charge $Q_{Z'}$ in Eq.~(\ref{YQZ}) for the fermion $\psi$.

\subsection{Heavy fermion effects at one loop}

The leading effects, due to the KK excitations that mix with the top
quark, can be computed using the results in
Refs.\cite{Lavoura:1992np,Bamert:1996px}.  The one-loop contributions
due to quarks to the $T$ and $S$ oblique parameters are
\beqa
T &=&\frac{3}{16\pi s^{2} c^{2} m^{2}_{Z}} \times
\nonumber \\
& &\left\{\sum_{i,j}
\left( V^{L}_{i j} V^{L*}_{i j} + V^{R}_{i j} 
V^{R*}_{i j} \right)
\theta_{+}(\MM_{i i}, \MM_{j j}) +
2 \,{\rm Re}\left( V^{L}_{i j} V^{R\ast}_{i j} \right)
\theta_{-}(\MM_{i i}, \MM_{j j}) \right.
\label{T:eq}
\\ & & \left. \hspace{3mm} \mbox{} -
\sum_{i} \sum^{i - 1}_{j}
\left( U^{L}_{i j} U^{L*}_{i j} + 
U^{R}_{i j} U^{R*}_{i j} \right)
\theta_{+}(\MM_{i i}, \MM_{j j}) +
2 \, {\rm Re}\left( U^{L}_{i j} U^{R\ast}_{i j} \right)
\theta_{-}(\MM_{i i}, \MM_{j j})
\right\}~,
\nonumber 
\label{S:eq} \\ [0.4em]
S &=& \frac{3}{4\pi}
\sum_{i,j} \left[
\left(
U_{ij}^L Y^L_{ji} +
U_{ij}^R Y^R_{ji} \right)
\bar{\chi}_{+}(\MM_{ii},\MM_{jj})
+ \left(
U_{ij}^L Y^R_{ji} +
U_{ij}^R Y^L_{ji} \right)
\bar{\chi}_{-}(\MM_{ii},\MM_{jj}) 
\right]~,
\nonumber \\ [-0.5em]
\eeqa
where the indices $i$, $j$ run over \text{all} fermions in
the theory (SM fermions and their KK excitations), 
\beqa
\theta_{+}(y_{1},y_{2}) &=& y^{2}_{1} + y^{2}_{2} - 
\frac{2 y^{2}_{1} y^{2}_{2}}{y^{2}_{1} - y^{2}_{2}} 
\ln \frac{y^{2}_{1}}{y^{2}_{2}}~,
\\
\theta_{-}(y_{1},y_{2}) &=& 2 y_{1} y_{2} 
\left( \frac{2 y^{2}_{1} y^{2}_{2}}{y^{2}_{1} - 
y^{2}_{2}} \ln \frac{y^{2}_{1}}{y^{2}_{2}} - 2 \right)~,
\eeqa
and
\beqa
\bar{\chi}_+(y_1,y_2) &=& \frac{5(y_1^4 + y_2^4) - 22 y_1^2 y_2^2}
{9 (y_1^2 - y_2^2)^2} +
\frac{3 y_1^2 y_2^2 (y_1^2 + y_2^2) - (y_1^6 + y_2^6)}{3 (y_1^2 -
y_2^2)^3}
\ln\left(\frac{y_1^2}{y_2^2}\right)
-\frac{2}{3} \ln \left(\frac{y_1 y_2}{\mu^2}\right)~,
\nonumber \\ [.7 em]
\bar{\chi}_{-}(y_1,y_2) &=&
\frac{y_1 y_2}{(y_1^2 - y_2^2)^3} \left[ y_1^4 - y_2^4 -
2 y_1^2 y_2^2 \ln \left(\frac{y_1^2}{y_2^2} \right) \right]~.
\eeqa
In the above, $\MM$ is the (diagonal) mass matrix, containing all
fermions in the theory, $V^{L}$ ($V^{R}$) is the matrix of couplings
of LH (RH) fermion fields to $W^{1}_{\mu}$ in the mass eigenstate
basis, and $U^{L}$ ($U^{R}$) is the corresponding matrix of couplings
to $W^{3}_{\mu}$.  The matrices $U^{L,R}$ are hermitian.  Finally,
$Y^{L,R}$ are the matrices of hypercharges for left- and right-handed
fermions in the mass eigenstate basis.

The leading one-loop contribution to the $Z b_L \bar{b}_L$ coupling,
that comes from the quarks with charge $2/3$, reads
\begin{eqnarray}
\delta g_{Zbb} =
\frac{\alpha}{2\pi}
&\bigg\{&
\sum_i \Big[ V^{L}_{i b} V^{L}_{i b}
\big(F_{SM}(r_i) + \tilde{F}(U^L_{ii}/2-1/2,U^R_{i
i}/2,r_i) \big) \Big] - F_{SM}(r_t)
\nonumber \\
&&+ \sum_{i<j}
 V^L_{i b} V^L_{j b} 
\mathcal{F}(U^L_{i j}/2,U^R_{ij},r_i,r_j)
\bigg\},
\label{Zbb:eq}
\end{eqnarray}
where $r_i\equiv m_i^2/m_W^2$ and
\beqa
F_{SM} (r)&=&
\frac{1}{8s^2} 
\frac{r(r-1)(r-6)+r(3r+2)\ln r}{(r-1)^2}, 
\nonumber
\\
\tilde{F}(\tilde{g}_L,\tilde{g}_R,r) &=&
\frac{1}{8 s^2} \left [
r \tilde{g}_L  \left(2 - \frac{4}{r-1} \ln r \right)
-r\tilde{g}_R  \left( \Delta +\frac{2r -5}{r-1} 
+\frac{r^2-2r+4}{(r-1)^2}\ln r \right)
\right], \nonumber
\\
{\mathcal F}(\tilde{g}_L,\tilde{g}_R,r,r^\prime) &=&
\frac{1}{4s^2(r^\prime-r)}\left \{ 2 \tilde{g}_L
\left[ 
\frac{r-1}{r^\prime-1} r^{\prime\,2} \ln r^\prime
-\frac{r^\prime-1}{r-1} r^2 \ln r \right]
\right. \nonumber  \\
&& \left .- \tilde{g}_R \sqrt{r r^\prime} \left [
(\Delta+1)(r^\prime-r)
+\frac{r^\prime+4}{r^\prime-1}r^\prime \ln r^\prime
-\frac{r+4}{r-1} r \ln r \right]
\right \}.
\eeqa
%



\begin{thebibliography}{99}  
  
\bibitem{Randall:1999ee}  
L.~Randall and R.~Sundrum,  
Phys.\ Rev.\ Lett.\  {\bf 83}, 3370 (1999)  
[arXiv:hep-ph/9905221].  

\bibitem{Grossman:1999ra}
  Y.~Grossman and M.~Neubert,
  Phys.\ Lett.\ B {\bf 474}, 361 (2000)
  [arXiv:hep-ph/9912408].

\bibitem{Gherghetta:2000qt}
  T.~Gherghetta and A.~Pomarol,
  Nucl.\ Phys.\ B {\bf 586}, 141 (2000)
  [arXiv:hep-ph/0003129];
  S.~J.~Huber and Q.~Shafi,
  Phys.\ Lett.\ B {\bf 498}, 256 (2001)
  [arXiv:hep-ph/0010195].

\bibitem{Peskin:1991sw}  
M.~E.~Peskin and T.~Takeuchi,  
Phys.\ Rev.\ D {\bf 46}, 381 (1992).  

\bibitem{Agashe:2003zs}
  K.~Agashe, A.~Delgado, M.~J.~May and R.~Sundrum,
  JHEP {\bf 0308}, 050 (2003)
  [arXiv:hep-ph/0308036].

\bibitem{Agashe:2006at}
  K.~Agashe, R.~Contino, L.~Da Rold and A.~Pomarol,
  Phys.\ Lett.\ B {\bf 641}, 62 (2006)
  [arXiv:hep-ph/0605341].
  
\bibitem{Djouadi:2006rk}
  A.~Djouadi, G.~Moreau and F.~Richard,
  arXiv:hep-ph/0610173.

\bibitem{Carena:2006bn}
  M.~Carena, E.~Pont\'{o}n, J.~Santiago and C.~E.~M.~Wagner,
  Nucl.\ Phys.\ B {\bf 759}, 202 (2006)
  [arXiv:hep-ph/0607106].
  
\bibitem{Cacciapaglia:2004rb}
  G.~Cacciapaglia, C.~Csaki, C.~Grojean and J.~Terning,
  Phys.\ Rev.\ D {\bf 71}, 035015 (2005)
  [arXiv:hep-ph/0409126].

\bibitem{gauge:Higgs:unification}
  N.~S.~Manton,
  Nucl.\ Phys.\ B {\bf 158}, 141 (1979);
  Y.~Hosotani,
  Phys.\ Lett.\ B {\bf 126}, 309 (1983);
  H.~Hatanaka, T.~Inami and C.~S.~Lim,
  %
  Mod.\ Phys.\ Lett.\ A {\bf 13}, 2601 (1998)
  [arXiv:hep-th/9805067];
  I.~Antoniadis, K.~Benakli and M.~Quiros,
  New J.\ Phys.\  {\bf 3}, 20 (2001)
  [arXiv:hep-th/0108005];
  M.~Kubo, C.~S.~Lim and H.~Yamashita,
  %
  Mod.\ Phys.\ Lett.\ A {\bf 17}, 2249 (2002)
  [arXiv:hep-ph/0111327];
  G.~von Gersdorff, N.~Irges and M.~Quiros,
  Nucl.\ Phys.\ B {\bf 635}, 127 (2002)
  [arXiv:hep-th/0204223];
  C.~Csaki, C.~Grojean and H.~Murayama,
  Phys.\ Rev.\ D {\bf 67}, 085012 (2003)
  [arXiv:hep-ph/0210133];
  N.~Haba, M.~Harada, Y.~Hosotani and Y.~Kawamura,
  Nucl.\ Phys.\ B {\bf 657}, 169 (2003)
  [Erratum-ibid.\ B {\bf 669}, 381 (2003)]
  [arXiv:hep-ph/0212035];
  C.~A.~Scrucca, M.~Serone and L.~Silvestrini,
  Nucl.\ Phys.\ B {\bf 669}, 128 (2003)
  [arXiv:hep-ph/0304220];
  C.~A.~Scrucca, M.~Serone, L.~Silvestrini and A.~Wulzer,
  JHEP {\bf 0402}, 049 (2004)
  [arXiv:hep-th/0312267];
  N.~Haba, Y.~Hosotani, Y.~Kawamura and T.~Yamashita,
  %
  Phys.\ Rev.\ D {\bf 70}, 015010 (2004)
  [arXiv:hep-ph/0401183];
  C.~Biggio and M.~Quiros,
  Nucl.\ Phys.\ B {\bf 703}, 199 (2004)
  [arXiv:hep-ph/0407348];
  Y.~Hosotani, S.~Noda and K.~Takenaga,
  %
  Phys.\ Lett.\ B {\bf 607}, 276 (2005)
  [arXiv:hep-ph/0410193];
  G.~Cacciapaglia, C.~Csaki and S.~C.~Park,
  %
  JHEP {\bf 0603}, 099 (2006)
  [arXiv:hep-ph/0510366];
  G.~Panico, M.~Serone and A.~Wulzer,
  Nucl.\ Phys.\ B {\bf 739}, 186 (2006)
  [arXiv:hep-ph/0510373];
  G.~Panico, M.~Serone and A.~Wulzer,
  Nucl.\ Phys.\ B {\bf 762}, 189 (2007)
  [arXiv:hep-ph/0605292];
  A.~Falkowski,
  arXiv:hep-ph/0610336.


\bibitem{Contino:2003ve}
  R.~Contino, Y.~Nomura and A.~Pomarol,
  Nucl.\ Phys.\ B {\bf 671}, 148 (2003)
  [arXiv:hep-ph/0306259].

\bibitem{Agashe:2004rs}
  K.~Agashe, R.~Contino and A.~Pomarol,
  Nucl.\ Phys.\ B {\bf 719}, 165 (2005)
  [arXiv:hep-ph/0412089];
  K.~Agashe and R.~Contino,
  Nucl.\ Phys.\ B {\bf 742}, 59 (2006)
  [arXiv:hep-ph/0510164];


\bibitem{Cacciapaglia:2006gp}
  G.~Cacciapaglia, C.~Csaki, G.~Marandella and J.~Terning,
  arXiv:hep-ph/0607146;

\bibitem{Buchmuller:1985jz}
  W.~Buchmuller and D.~Wyler,
  Nucl.\ Phys.\ B {\bf 268}, 621 (1986).


\bibitem{Han:2004az}
  Z.~Han and W.~Skiba,
  Phys.\ Rev.\ D {\bf 71}, 075009 (2005)
  [arXiv:hep-ph/0412166];
  Z.~Han,
  Phys.\ Rev.\ D {\bf 73}, 015005 (2006)
  [arXiv:hep-ph/0510125].


\bibitem{delAguila:2000aa}
  F.~del Aguila, M.~Perez-Victoria and J.~Santiago,
  Phys.\ Lett.\ B {\bf 492}, 98 (2000)
  [arXiv:hep-ph/0007160];
  JHEP {\bf 0009}, 011 (2000)
  [arXiv:hep-ph/0007316].


\bibitem{delAguila:2000kb}
  F.~del Aguila and J.~Santiago,
  Phys.\ Lett.\ B {\bf 493}, 175 (2000)
  [arXiv:hep-ph/0008143];
  arXiv:hep-ph/0011143.

\bibitem{DelAguila:2001pu}
  F.~Del Aguila and J.~Santiago,
  JHEP {\bf 0203}, 010 (2002)
  [arXiv:hep-ph/0111047].


\bibitem{Lavoura:1992np}
  L.~Lavoura and J.~P.~Silva,
  Phys.\ Rev.\ D {\bf 47}, 2046 (1993).

\bibitem{Bamert:1996px}
  P.~Bamert, C.~P.~Burgess, J.~M.~Cline, D.~London and E.~Nardi,
  Phys.\ Rev.\ D {\bf 54}, 4275 (1996)
  [arXiv:hep-ph/9602438].

\bibitem{Yao:2006px}
  W.~M.~Yao {\it et al.}  [Particle Data Group],
  J.\ Phys.\ G {\bf 33}, 1 (2006).

\bibitem{gaugeKK}
  K.~Agashe, A.~Belyaev, T.~Krupovnickas, G.~Perez and J.~Virzi,
  arXiv:hep-ph/0612015.

\bibitem{Agashe:2004bm}
  K.~Agashe and G.~Servant,
  JCAP {\bf 0502}, 002 (2005)
  [arXiv:hep-ph/0411254].

\bibitem{Hirn:2006nt}
  J.~Hirn and V.~Sanz,
  Phys.\ Rev.\ Lett.\  {\bf 97}, 121803 (2006)
  [arXiv:hep-ph/0606086];
  arXiv:hep-ph/0612239.

\bibitem{Agashe:2006wa}
  K.~Agashe, G.~Perez and A.~Soni,
  arXiv:hep-ph/0606293.

\bibitem{Aquino:2006vp}
  P.~M.~Aquino, G.~Burdman and O.~J.~P.~Eboli,
  arXiv:hep-ph/0612055.

\bibitem{Wj:note}
``Search for Heavy Top $t^\prime \to W q$  in Lepton Plus Jets
Events'',
CDF-note-8495 and J. Conway, \textit{ private communication}.

\bibitem{Zj:note} 
``Search for New Particles Decaying to $Z^0+$jets'',
CDF-note-8590.

\bibitem{Carena:2002es}
  M.~Carena and H.~E.~Haber,
  Prog.\ Part.\ Nucl.\ Phys.\  {\bf 50}, 63 (2003)
  [arXiv:hep-ph/0208209].
  
\bibitem{Aguilar-Saavedra:2005pv}
  J.~A.~Aguilar-Saavedra,
  Phys.\ Lett.\ B {\bf 625}, 234 (2005)
  [Erratum-ibid.\ B {\bf 633}, 792 (2006)]
  [arXiv:hep-ph/0506187];
  J.~A.~Aguilar-Saavedra,
  PoS {\bf TOP2006}, 003 (2006)
  [arXiv:hep-ph/0603199].

\bibitem{T:littleHiggs}
D.~Costanzo, ATL-PHYS-2004-004;
  G.~Azuelos {\it et al.},
  Eur.\ Phys.\ J.\ C {\bf 39S2}, 13 (2005)
  [arXiv:hep-ph/0402037].

\bibitem{Aguilar-Saavedra:2006gw}
  J.~A.~Aguilar-Saavedra,
  arXiv:hep-ph/0603200.
  
\bibitem{Cacciapaglia:2006mz}
  G.~Cacciapaglia, C.~Csaki, G.~Marandella and J.~Terning,
  arXiv:hep-ph/0611358.

\bibitem{Contino:2006qr}
  R.~Contino, L.~Da Rold and A.~Pomarol,
  arXiv:hep-ph/0612048.

  
\end{thebibliography}
\end{document}